\documentstyle[12pt,a4wide,cite]{article}
\renewcommand{\arraystretch}{1.4}

\newcommand{\CQ}{\left( {\langle\bar q q \rangle^{(2)}(\mu)\over f^2}\right)}

\newcommand{\ds}{\displaystyle}
\renewcommand{\a}{\alpha}

\newcommand{\g}{\gamma}
\newcommand{\s}{\sigma}

\newcommand{\eps}{\varepsilon}

\newcommand{\as}{\alpha_{s}}
\newcommand{\bea}{\begin{eqnarray}}
\newcommand{\eea}{\end{eqnarray}}
\newcommand{\beq}{\begin{equation}}
\newcommand{\eeq}{\end{equation}}
\newcommand{\nn}{\nonumber}
\newcommand{\fr}{\frac}

\newcommand{\hl}{\hline}

\newcommand{\real}{{\rm Re}}

\newcommand{\pr}{Phys.~Rev. }
\newcommand{\pl}{Phys.~Lett. }
\newcommand{\np}{Nucl.~Phys. }
\newcommand{\prl}{Phys.~Rev.~Lett. }
\newcommand{\cO}{{\cal O}}
\newcommand{\cL}{{\cal L}}
\newcommand{\cQ}{{\cal Q}}
\newcommand{\cA}{{\cal A}}

\newcommand{\lsim}{~{}_{\textstyle\sim}^{\textstyle <}~}
\newcommand{\ba}{\begin{array}{c}}
\newcommand{\bat}{\begin{array}{cc}}
\newcommand{\ea}{\end{array}}
\def\eqn#1{(\ref{#1})}
\def\slashchar#1{\setbox0=\hbox{$#1$}\dimen0=\wd0%
\setbox1=\hbox{/}\dimen1=\wd1%
\ifdim\dimen0>\dimen1%
\rlap{\hbox to
\dimen0{\hfil/\hfil}}#1\else                                     
\rlap{\hbox to \dimen1{\hfil$#1$\hfil}}/\fi}
\begin{document}

\begin{titlepage}
\begin{flushright}
{IFIC/00-31 \\ FTUV/01-0502 \\ SISSA 37/2001/EP }
\end{flushright}
\vskip 2cm
\centerline{\LARGE \bf
The Standard Model Prediction for  $\eps '  /\eps$
}

\vskip 1cm
\centerline{E. Pallante}
\vskip 0.5cm
\centerline{SISSA, Via Beirut 2-4, I-34013 Trieste, Italy}

\vskip 1cm
\centerline{A. Pich,\ I. Scimemi}
\vskip 0.5cm
\centerline{Departament de F\'{\i}sica Te\`orica, IFIC,
CSIC --- Universitat de Valencia}
\centerline{ Edifici d'Instituts de Paterna,
Apt. Correus 22085, E--46071, Val\`encia,  Spain}

\vskip 2.5cm

\begin{abstract}
We present a detailed analysis of $\eps'/\eps$ within the
Standard Model, taking into account the strong enhancement
through final--state interactions  identified
in refs.~\cite{PP:00} and \cite{PP:00b}.
The relevant hadronic matrix elements are fixed at leading order
in the $1/N_C$ expansion, through a matching procedure between
the effective short--distance Lagrangian and its
corresponding low--energy description in Chiral
Perturbation Theory. All large logarithms are summed up, both at
short and long distances.
Two different numerical analyses are performed, using either
the experimental or the theoretical value for $\eps$,
with compatible results. We obtain
${\rm Re}\left(\eps'/\eps\right) = (1.7 \pm 0.9)  \cdot  10^{-3} \ .$
The error is dominated by the uncertainty in the value of the
strange quark mass and the estimated corrections from unknown
$1/N_C$--suppressed local contributions.
A better   estimate of the strange quark mass would reduce
the uncertainty to about 30$\%$.
The Standard Model prediction agrees  with the present
experimental world average
$  { \rm Re}\left(\eps'/\eps\right) =(1.93 \pm 0.24) \cdot 10^{-3}\ $.
\end{abstract}

\end{titlepage} \newpage

\section{Introduction}
\label{sec:introd}

The CP--violating ratio  $\eps '/\eps $  constitutes a fundamental
test for our under\-stan\-ding of fla\-vour--chan\-ging phenomena within
the Standard Model framework.
It represents a great source of inspiration for physics research
and has motivated in recent years a very interesting scientific
controversy, both on the experimental and theoretical sides.

The experimental status has been clarified recently.
The CERN NA48 collaboration~\cite{na48} has announced a
preliminary value
\beq
{ \rm Re} \left(\eps '/\eps\right)=(1.40 \pm 0.43) \cdot 10^{-3} \ .
\eeq
A larger result was obtained by the
Fermilab KTeV collaboration~\cite{ktev},
\beq
{ \rm Re} \left(\eps '/\eps\right) = (2.80 \pm 0.41) \cdot 10^{-3} \ .
\eeq
The present world average~\cite{na48},
\beq
{ \rm Re} \left(\eps '/\eps\right) =(1.93 \pm 0.24) \cdot 10^{-3} \ ,
\eeq
provides clear evidence for a non-zero value and,
therefore, direct CP violation phenomena.

The theoretical status is more involved and not very satisfactory.
There is no universal
agreement on the $\eps '/\eps$ value predicted by the
Standard Model, since different groups, using different models or
approximations, obtain different results
\cite{munich,rome,trieste,FA:00,dortmund,BP:00,we:00,dubna,taipei,NA:00}.
Nevertheless, it has been often claimed that the
Standard Model predicts a too small value of $\eps '/\eps$, failing
to reproduce its experimental world average by at least a factor of two.
This claim has generated a very intense theoretical activity,
searching for new sources of CP violation beyond the Standard Model
framework \cite{beyond}.

It has been pointed out \cite{PP:00} that the theoretical
short--distance evaluations of $\eps '/\eps$ had overlooked the
important role of final--state interactions (FSI) in $K\to\pi\pi$ decays.
Although it has been known for more than a decade that the rescattering
of the two final pions induces a large correction to the isospin--zero
decay amplitude, this effect was not taken properly into account in
the theoretical predictions. From the measured 
$\pi$--$\pi$ phase shifts one can easily infer that
FSI generate a strong enhancement of the
predicted $\eps '/\eps$ value, by roughly the needed factor
of two \cite{PP:00,PP:00b}.
A detailed analysis of the corrections induced by FSI
has been already given in ref.~\cite{PP:00b}, where the low-energy
(infrared) physics involved has been investigated and the size of the
FSI enhancement and the associated uncertainties have been quantified.

In this paper, we present a complete reevaluation of $\eps '/\eps$
within the Standard Model. We will show that with our present understanding
of the different inputs, it is possible to pin down the prediction of
this important parameter with a theoretical accuracy of about 50\%.
In order to
achieve this goal, one needs to identify the most important corrections
and find appropriate expansion parameters to perform a
perturbative approach with well--defined power counting.

The large--$N_C$ expansion \cite{tHO:74,WI:79},
with $N_C$ the number of QCD colours,
turns out to be a very useful tool to organize the calculation. It is
a unique non-perturbative approach, with a clear meaning within the
usual perturbative expansion in powers of the QCD coupling.
At leading (non-trivial) order in $1/N_C$ it is possible to compute
all needed ingredients and, what is even more important, the matching
between short-- and long--distance physics can be done exactly.
Moreover, FSI are zero at leading order in $1/N_C$; this allows a clear
separation of these corrections, avoiding any possible ambiguity or
double--counting.

Since $N_C=3$ in the real world, the natural size to be expected
for the $1/N_C$--suppressed contributions is 30\%. 
Actually, there is a quite compelling
phenomenological evidence that those corrections are usually smaller.
For this to be true, however, one needs to make sure that the $1/N_C$
expansion does not involve large logarithms \cite{BBG87}; 
i.e. one should expand
in powers of $1/N_C$ and not in powers of $\frac{1}{N_C} \ln{(M/m)}$, with
$M\gg m$ two widely separated scales.
Large logarithms are in fact the main source of complications
in low--energy flavour--changing processes,
because the electroweak scale $M_W$
where the short--distance quark transition takes place is much larger
than the long--distance hadronic scale.

The large short--distance logarithms can be summed up with the use of the
Operator Product Expansion (OPE) \cite{WI:69}
and the renormalization group \cite{RGroup}. The proper
way to proceed makes use of modern Effective Field Theory techniques
\cite{EFT}.
One starts above the electroweak scale where the flavour--changing process,
in terms of quarks, leptons and gauge bosons, can be analyzed
within the usual gauge--coupling perturbative expansion in a rather
straightforward way. The renormalization group is used to evolve down
in energy  from the scale $M_Z$, where the top quark and the $Z$ and
$W^\pm$ bosons are integrated out. That means that one changes to a
different Effective Theory where those heavy particles are no longer
explicit degrees of freedom. The new Lagrangian contains a tower of
operators constructed with the light fields only, which scale as
powers of $1/M_Z$. The information on the heavy fields is hidden in
the (Wilson) coefficients of those operators, which are fixed by
``matching'' the high-- and low--energy theories at the point $\mu=M_Z$.
One follows the evolution further to lower energies, using the
Effective Theory renormalization group equations, until a new particle
threshold is encountered. Then, the whole procedure of integrating the
new heavy scale and matching to another Effective Field Theory starts
again. In this way, one proceeds down to scales $\mu < m_c$.

In this picture, the physics is described by a chain of different
Effective Field Theories, with different particle content, which match
each other at the corresponding boundary (heavy threshold). This
procedure permits to perform an explicit summation of large logarithms
$t\equiv\ln{(M/m)}$, where $M$ and $m$ refer to any scales
appearing in the evolution.
One gets finally an effective $\Delta S=1$ Lagrangian, defined in the
three--flavour theory \cite{GLAM:74,VZS:75,GW:79,BU:99},
\beq\label{eq:Leff}
 {\cal L}_{\mathrm eff}^{\Delta S=1}= - \frac{G_F}{\sqrt{2}}
 V_{ud}^{\phantom{*}}\,V^*_{us}\,  \sum_i  C_i(\mu) \; Q_i (\mu) \; ,
 \label{eq:lag}
\eeq
which is a sum of local four--fermion operators $Q_i$,
constructed with the light degrees of freedom,
\bea
\begin{array}{rcl}
Q_{1} & = & \left( \overline{s}_{\alpha} u_{\beta}  \right)_{\rm V-A}
            \left( \overline{u}_{\beta}  d_{\alpha} \right)_{\rm V-A}
\, , \\[1ex]
Q_{2} & = & \left( \overline{s} u \right)_{\rm V-A}
            \left( \overline{u} d \right)_{\rm V-A}
\, , \\[1ex]
Q_{3,5} & = & \left( \overline{s} d \right)_{\rm V-A}
   \sum_{q} \left( \overline{q} q \right)_{\rm V\mp A}
\, , \\[1ex]
Q_{4,6} & = & \left( \overline{s}_{\alpha} d_{\beta}  \right)_{\rm V-A}
   \sum_{q} ( \overline{q}_{\beta}  q_{\alpha} )_{\rm V\mp A}
\, , \\[1ex]
Q_{7,9} & = & \frac{3}{2} \left( \overline{s} d \right)_{\rm V-A}
         \sum_{q} e_q \,\left( \overline{q} q \right)_{\rm V\pm A}
\, , \\[1ex]
Q_{8,10} & = & \frac{3}{2} \left( \overline{s}_{\alpha}
                                    d_{\beta} \right)_{\rm V-A}
  \sum_{q} e_q \,\left( \overline{q}_{\beta} q_{\alpha}\right)_{\rm V\pm A}
\, ,
\end{array} 
\label{eq:ope}
\eea
modulated
by Wilson coefficients $C_i(\mu)$ which are functions of the
heavy masses.
Here $\alpha$, $\beta$ denote colour indices
and $e_q$ are the quark charges
($e_u = 2/3$,  $e_d=e_s =-1/3$). Colour
indices for the colour singlet operators are omitted.
The labels \mbox{$(V\pm A)$} refer to the Dirac structures
\mbox{$\gamma_{\mu} (1 \pm \gamma_5)$}.

We have explicitly factored out the Fermi coupling $G_F$
and the Cabibbo--Kobayashi--Maskawa (CKM) matrix elements
$V_{ij}$ containing the usual Cabibbo suppression of $K$ decays.
The unitarity of the CKM matrix allows to write the Wilson coefficients
in the form
\beq
C_i(\mu) =  z_i(\mu) + \tau\ y_i(\mu) \; ,
\label{eq:Lqcoef}
\eeq
where
$\tau = - V_{td}^{\phantom{*}} V_{ts}^{*}/V_{ud}^{\phantom{*}} V_{us}^{*}$.
The CP--violating
decay amplitudes  are proportional to the  $y_i$ components.

The overall renormalization scale $\mu$ separates
the short-- ($M>\mu$) and long-- ($m<\mu$) distance contributions,
which are contained in $C_i(\mu)$ and $Q_i$, respectively.
The physical amplitudes are independent of $\mu$; thus, the
explicit scale (and scheme) dependence of the Wilson coefficients should
cancel exactly with the corresponding dependence of the $Q_i$
matrix elements between on-shell states.

Our knowledge of $\Delta S=1$ transitions has improved qualitatively
in recent years, thanks to the completion of the next-to-leading
logarithmic--order calculation of the Wilson coefficients
\cite{buras1,ciuc1}.
All gluonic corrections of $\cO(\as^n t^n)$ and $\cO(\as^{n+1} t^n)$
are already known. Moreover the full $m_t/M_W$ dependence (to first order
in $\a_s$ and $\a$) has been taken into account at the electroweak scale.
We will fully use this information up to scales $\mu\sim \cO(1\; {\rm GeV})$,
without making any unnecessary expansion in powers of $1/N_C$.

{\renewcommand{\arraystretch}{1.0}
\begin{figure}[tbh]    
\setlength{\unitlength}{0.75mm} \centering
\begin{picture}(165,120)
\put(0,0){\makebox(165,120){}}
\thicklines
\put(10,105){\makebox(40,15){\large Energy Scale}}
\put(58,105){\makebox(36,15){\large Fields}}
\put(110,105){\makebox(40,15){\large Effective Theory}}
\put(8,108){\line(1,0){149}} {\large
\put(10,75){\makebox(40,27){$M_W$}}
\put(58,75){\framebox(36,27){$\ba W, Z, \gamma, g \\
     \tau, \mu, e, \nu_i \\ t, b, c, s, d, u \ea $}}
\put(110,75){\makebox(40,27){Standard Model}}

\put(10,40){\makebox(40,18){$\lsim m_c$}}
\put(58,40){\framebox(36,18){$\ba  \gamma, g  \; ;\; \mu ,  e, \nu_i
             \\ s, d, u \ea $}}
\put(110,40){\makebox(40,18){$\cL_{\mathrm{QCD}}^{(n_f=3)}$, \
             $\cL_{\mathrm{eff}}^{\Delta S=1,2}$}}

\put(10,5){\makebox(40,18){$M_K$}}
\put(58,5){\framebox(36,18){$\ba\gamma \; ;\; \mu , e, \nu_i  \\
            \pi, K,\eta  \ea $}}
\put(110,5){\makebox(40,18){$\chi$PT}}
\linethickness{0.3mm}
\put(76,37){\vector(0,-1){11}}
\put(76,72){\vector(0,-1){11}}
\put(80,64.5){OPE}
\put(80,29.5){$N_C\to\infty$}}
\end{picture}
\caption{Evolution from $M_W$ to $M_K$.  
  \label{fig:eff_th}}
\end{figure}
}

In order to predict physical amplitudes one is still
confronted with the calculation of hadronic matrix elements of
quark operators. This is a very difficult problem, which so far
remains unsolved.
As indicated in figure~\ref{fig:eff_th}, below the resonance region
one can use global symmetry considerations to define another
Effective Field Theory in terms of the QCD Goldstone bosons
($\pi$, $K$, $\eta$). The Chiral Perturbation Theory ($\chi$PT)
formulation of the Standard Model \cite{WE:79,GL:85,EC:95,PI:95,ME:93}
is an ideal framework to describe
the pseudoscalar--octet dynamics, through a perturbative expansion
in powers of momenta and light quark masses over the chiral symmetry
breaking scale ($\Lambda_\chi\sim 1\; {\rm GeV}$).
Chiral symmetry fixes the allowed $\chi$PT operators,
at a given order in momenta. The only remaining
problem is then the calculation of the corresponding chiral couplings
from the effective short--distance Lagrangian; this requires
to perform the matching between the two Effective Field Theories.

It is here where the $1/N_C$ expansion proves to be useful.
At leading order in $1/N_C$, the matching
between the 3--flavour quark theory and $\chi$PT can be done exactly.
We will determine the needed chiral couplings
in the large--$N_C$ limit, in a quite straightforward way.
The scale and scheme dependences of the short--distance Wilson
coefficients are of course completely removed in the matching process,
at leading order in $1/N_C$. Any remaining dependences are higher--order
in the $1/N_C$ expansion and, thus, numerically suppressed; they are
 included in our estimated theoretical uncertainty.

There is still an important source of large logarithms that needs to be
identified and kept under control. The FSI of the pseudo-Goldstone pions
generate large infrared logarithms, involving the light pion mass,
which are next-to-leading in $1/N_C$. These chiral logarithms can be
computed within the effective $\chi$PT framework. Moreover, as shown in
refs.~\cite{PP:00,PP:00b} they can be exponentiated to all orders in
the momentum expansion. Since this is a $1/N_C$ suppressed
(but numerically large) effect, it generates an important correction,
not included in the previous leading--order determination of chiral
couplings. 

The paper is organized as follows. The usual isospin formalism for
$K\to\pi\pi$ decays and the relevant formulae for $\eps'/\eps$ are
collected in section~\ref{sec:isospin}.
Section~\ref{sec:chpt} presents the low--energy $\chi$PT description.
The matching between the short-- and long--distance effective theories is
performed in section~\ref{sec:matching}, at leading order in $1/N_C$.
Section~\ref{sec:N_C_results} summarizes the large--$N_C$ predictions for
the different isospin amplitudes.
The one-loop chiral corrections are discussed in 
section~\ref{sec:loops}.
Section~\ref{sec:FSI} incorporates higher--order corrections induced by
FSI, within the chiral framework.
The Standard Model prediction for $\eps'/\eps$ is worked out in
section~\ref{sec:num}, where two different numerical analyses are presented.
The first one incorporates the experimental value of $\eps$, while in
the second one its theoretical prediction is used instead. Both analyses
give compatible results. Our conclusions are finally given in
section~\ref{sec:summary}.
We have collected in several appendices
the analytical results from
the one-loop chiral calculation of the different
$K\to\pi\pi$ amplitudes.

\section{$K\to \pi\pi$ Amplitudes}
\label{sec:isospin}

We  adopt the usual isospin decomposition:
\bea\label{eq:AI}
A[K^0\to \pi^+\pi^-] &\equiv&  \cA_0 + {1\over \sqrt{2}}\,  \cA_2
\, , \nonumber\\
A[K^0\to \pi^0\pi^0] &\equiv&  \cA_0 - \sqrt{2}\,  \cA_2\, .
\eea
The complete amplitudes
${\cal A}_I \equiv A_I\, \exp{\left\{i\delta_0^I\right\}}$
include the strong phase shifts $\delta_0^I$.
The S--wave $\pi$-$\pi$ scattering generates a large
phase-shift difference between the $I=0$ and $I=2$ partial
waves \cite{GM:91}:
\beq
\left(\delta_0^0 - \delta_0^2\right)(M_K^2) = 45^\circ\pm 6^\circ
\, .
\eeq
There is a corresponding dispersive FSI effect in 
the moduli of the isospin amplitudes, because the real and 
imaginary parts are related by analyticity and unitarity.
The presence of such a large phase-shift difference clearly signals
an important FSI contribution to $A_I$.

In terms of the $K\to\pi\pi$ isospin amplitudes,
\beq
{\varepsilon^\prime\over\varepsilon} =
\; e^{i\Phi}\; {\omega\over \sqrt{2}\vert\eps\vert}\;\left[
{\mbox{Im}(A_2)\over\mbox{Re} (A_2)} - 
{\mbox{Im}(A_0)\over \mbox{Re} (A_0)}
 \right] \, .
\eeq
Owing to the well-known ``$\Delta I=1/2$ rule'', $\eps'/\eps$ is
suppressed by the ratio
\beq
\omega = \mbox{Re} (A_2)/\mbox{Re} (A_0) \approx 1/22\, .
\eeq
The phases of $\eps'$ and $\eps$ turn out to be nearly equal:
\beq
\Phi \approx \delta^2_0-\delta^0_0+\frac{\pi}{4}\approx 0 \, .
\eeq
The CP--conserving amplitudes $\mbox{Re} (A_I)$, their ratio
$\omega$ and $|\eps|$ are usually set to their experimentally
determined values. A theoretical calculation is then only needed
for $\mbox{Im} (A_I)$.

Using the short--distance Lagrangian \eqn{eq:Leff}, the CP--violating ratio
$\eps'/\eps$ can be written as \cite{munich}
\beq
{\varepsilon^\prime\over\varepsilon} \, = \,
\mbox{Im}\left(V_{ts}^* V_{td}^{\phantom{*}}\right)\, e^{i\Phi}\; 
{G_F\over 2 \vert\varepsilon\vert}\; {\omega\over |\mbox{Re}(A_0)|}\;
\left [P^{(0)}\, (1-\Omega_{IB}) - {1\over \omega} \,P^{(2)}\right ]\, ,
\label{EPS}
\eeq
where the quantities
\beq
P^{(I)}= \sum_i\, y_i(\mu)\;
\langle (\pi\pi )_I\vert Q_i\vert K\rangle 
\eeq
contain the contributions from hadronic matrix elements 
with isospin $I$ \ and
\beq
\Omega_{IB} = {1\over \omega}\,
{\mbox{Im}(A_2)_{\mbox{\small{IB}}}  \over \mbox{Im}(A_0)}
\label{eq:isospin}
\eeq
parameterizes isospin breaking corrections.
The factor $1/\omega$ enhances the relative weight of the
$I=2$ contributions.

The hadronic matrix elements 
$\langle (\pi\pi )_I\vert Q_i\vert K\rangle $
are usually parameterized
in terms of the so-called bag parameters $B_i$, which measure them
in units of their vacuum insertion approximation values.
In the Standard Model, $P^{(0)}$ and $P^{(2)}$ turn out to
be dominated by the contributions from 
the QCD penguin operator $Q_6$ and
the electroweak penguin operator $Q_8$, respectively \cite{trieste}.
Thus, to a very good approximation,
$\varepsilon'/\varepsilon$ can be written (up to global factors)
as \cite{munich}
\beq
{\varepsilon'\over\varepsilon} \,\sim\,
\left [ B_6^{(1/2)}(1-\Omega_{IB}) - 0.4 \, B_8^{(3/2)}
 \right ]\, .
\label{EPSNUM}
\eeq

The isospin--breaking correction coming from $\pi^0$-$\eta$  mixing
was originally estimated to be $\Omega_{IB}^{\pi^0\eta}=0.25$ 
\cite{Omega,BG:87}. Together with the usual ansatz $B_i\sim 1$, 
this produces a large numerical cancellation
in eq.~\eqn{EPSNUM} leading to low values of $\eps'/\eps$
around $7\cdot 10^{-4}$.
A recent improved calculation of $\pi^0$-$\eta$  mixing
at $\cO(p^4)$ in $\chi$PT has found the result \cite{EMNP:00}
\beq\label{eq:omIB}
\Omega_{IB}^{\pi^0\eta}\, =\, 0.16\pm 0.03 \, .
\eeq
This smaller number, slightly increases the naive estimate of $\eps'/\eps$.

\section{Chiral Perturbation Theory Description}
\label{sec:chpt}

In the limit $m_u$, $m_d$, $m_s \rightarrow 0$, the QCD Lagrangian for
light quarks has a $SU(3)_L \otimes SU(3)_R$ symmetry,
which is spontaneously broken to $SU(3)_V$.
The lightest particles of the hadronic spectrum, the pseudoscalar
octet ($\pi$, $K$, $\eta$), can be identified with the
corresponding Goldstone bosons.
Their low--energy interactions can be analyzed within $\chi$PT
\cite{WE:79,GL:85,EC:95,PI:95,ME:93}, which is an expansion in
terms of momenta and meson (quark) masses.
The Goldstone fields are parameterized as
\beq\label{pi}
{\bf \Phi}= \left ( \begin{array}{ccc}
\sqrt{\frac{1}{2}} \pi^0 + \sqrt{\frac{1}{6}} \eta & \pi^+ & K^+ \\
\pi^- & - \sqrt{\frac{1}{2}} \pi^0 +\sqrt{\frac{1}{6}} \eta & K^0 \\
K^- & \bar{K}^0 & -\sqrt{\frac{2}{3}} \eta \end{array}
\right ),
\eeq
and appear in the Lagrangian via the exponential representation
$U=\exp(\sqrt{2} i {\bf\Phi} /f)$,  with $f \sim f_\pi = 92.4$
MeV\ the pion decay constant at lowest order.
Under a chiral transformation
$g\equiv (g_L,\, g_R) \in SU(3)_L \otimes SU(3)_R$,
the matrix $U$ changes as \
$U\;\rightarrow\; g_R\, U\, g_L^\dagger \, . $

The effect of strangeness--changing non-leptonic
weak interactions with $\Delta S=1$ is incorporated \cite{CR:67}
in the low--energy chiral theory as a perturbation to
the strong effective Lagrangian. At lowest order,
the most general effective bosonic Lagrangian, with the same
$SU(3)_L\otimes SU(3)_R$ transformation properties and quantum numbers
as the short--distance Lagrangian \eqn{eq:Leff}, contains three terms:
\bea\label{eq:lg8_g27}
\cL_2^{\Delta S=1} &=& -{G_F \over \sqrt{2}}\,  V_{ud}^{\phantom{*}} V_{us}^*
\; f^4\;\Bigg\{ g_8 \;\left[ \langle\lambda L_{\mu} L^{\mu}\rangle 
+ e^2 f^2 g_{ew}\;\langle\lambda U^\dagger \cQ U\rangle \right]
\Biggr. \nn\\ &&\qquad\qquad\qquad\quad\;\mbox{}  +
g_{27}\,\left( L_{\mu 23} L^\mu_{11} + {2\over 3} L_{\mu 21} L^\mu_{13}
\right)\Bigg\} \, ,
\eea
where the matrix $L_{\mu}=-i U^\dagger D_\mu U$  represents the octet of
$V-A$ currents, at lowest order in derivatives,
$\cQ= {\rm diag}(\frac{2}{3},-\frac{1}{3},-\frac{1}{3})$ is the quark
charge matrix,
$\lambda\equiv (\lambda^6 - i \lambda^7)/2$ projects onto the
$\bar s\to \bar d$ transition [$\lambda_{ij} = \delta_{i3}\delta_{j2}$]
and $\langle {\mbox{A}}\rangle$ denotes the flavour trace of A.

The chiral couplings $g_8$ and $g_{27}$ measure the strength of the two
parts of the effective Lagrangian \eqn{eq:Leff} transforming as
$(8_L,1_R)$ and $(27_L,1_R)$, respectively, under chiral rotations.
Chiral symmetry forces the lowest--order Lagrangian to contain at least
two derivatives (Goldstone bosons are free particles at zero momenta).
In the presence of electroweak interactions, however, the explicit breaking
of chiral symmetry generated by the quark charge matrix $\cQ$ induces
the $\cO(p^0)$ operator $\langle\lambda U^\dagger \cQ U\rangle$
\cite{BW:84,GRW:86}, transforming as $(8_L,8_R)$ under the chiral group.
In the usual chiral counting $e^2\sim \cO(p^2)$ and, therefore, the
$g_{ew}$ term appears at the same order in the derivative expansion than
$g_8$ and $g_{27}$.
One additional term \cite{BE:85} proportional to the quark mass matrix,
which transforms as $(8_L,1_R)$, has not been written in the lowest--order
Lagrangian (\ref{eq:lg8_g27}),
since it does not contribute\footnote{
The contributions of this term to $K\to \pi\pi$ amplitudes
vanish at $\cO(p^2)$, while at $\cO(p^4)$ they can be reabsorbed
through a redefinition of the local $\cO(p^4)$ $\Delta S=1$
chiral couplings \cite{CR:86,KA91,BPP}.}
to physical $K\to \pi\pi$ matrix elements
\cite{CR:86,KA91,BPP}.

The tree--level $K\to\pi\pi$ amplitudes generated
by the $\cO(p^2)$ $\chi$PT Lagrangian (\ref{eq:lg8_g27}) are:
\bea
\label{eq:AMP_2}
\cA_0 &\! =&\! -{G_F\over \sqrt{2}} \, V_{ud}^{\phantom{*}} V_{us}^*\;\sqrt{2}\,
f\;\left\{\left ( g_8+{1\over 9}\, g_{27}\right )(M_K^2-M_\pi^2)
-{2\over 3}\, f^2\, e^2 \; g_8\; g_{ew}\right\}\; ,
\nonumber\\
\cA_2&\! =&\!  -{G_F\over \sqrt{2}} \, V_{ud}^{\phantom{*}} V_{us}^*\; {2\over 9}
\,f\;\biggl\{ 5\, g_{27}\, (M_K^2-M_\pi^2) -3\, f^2\, e^2\; g_8\; g_{ew}
\biggr\}
\; .
\eea
The strong phase shifts are zero at lowest order.
Taking the measured phase shifts into account,
the moduli of $g_8$ and $g_{27}$ can be extracted from
the CP--conserving $K \rightarrow 2 \pi$ decay rates.
A lowest--order phenomenological analysis \cite{PGR:86}, 
neglecting\footnote{
A general analysis of isospin breaking and electromagnetic corrections to
$K\to\pi\pi$ transitions is presented in 
refs.~\protect\cite{EIMNP:00,CDG:99,CG:00}.
} 
the tiny electroweak corrections proportional to $e^2 g_{ew}$, gives:
\beq\label{eq:g8_g27}
\left| g_8 \right| \simeq 5.1\, , \qquad\qquad
\left| g_{27} \right| \simeq 0.29 \, .
\eeq
The huge difference between these two couplings
shows the well--known
enhancement of octet $\vert\Delta I\vert = 1/2$ transitions.

The isospin amplitudes $\cA_I$ have been
computed up to next--to--leading order in the chiral expansion
\cite{KA91,BPP,EIMNP:00,CDG:99,CG:00,EKW:93}. 
Decomposing the isoscalar amplitudes in their octet and 27--plet components as
${ \cA}_0 = { \cA}_0^{(8)}+{ \cA}_0^{(27)}$, 
the results of those calculations can be written in the form:
\bea 
\label{ONELOOP_8} 
 \cA_0^{(8)} \, =\,
-{G_F\over \sqrt{2}} V^{\phantom{\ast}}_{ud}V^\ast_{us} 
\,\sqrt{2}\, f_\pi \; g_8
&&\!\!\!\!\!\!\! \Biggl\{ (M_K^2-M_\pi^2)\,
 \left[1+\Delta_L{\cal A}_0^{(8)} +\Delta_C{\cal A}_0^{(8)}
 \right] \Biggr. \nn \\
&& \!\!\!\! \Biggl.\mbox{}
-{2\over 3}\, e^2\, f^2_\pi\,
\left[  g_{ew}\,\left( 1 +\Delta_L{\cal A}_{0}^{(ew)}\right) +
\Delta_C{\cal A}_{0}^{(ew)}
\right] \Biggr\}\quad 
\eea
for the octet isoscalar amplitude,   
\beq 
\label{ONELOOP_27} 
 \cA_0^{(27)} \, =\,
-{G_F\over \sqrt{2}} V^{\phantom{\ast}}_{ud}V^\ast_{us} 
\, {\sqrt{2}\over 9}\, f_\pi\, g_{27}\, (M_K^2-M_\pi^2)\,\left[ 
 1+\Delta_L{\cal A}_0^{(27)} +\Delta_C{\cal A}_0^{(27)} \right]
 \,
\eeq 
for the 27--plet isoscalar amplitude and 
\bea 
\label{ONELOOP_2} 
 \cA_2 \, =\,
-{G_F\over \sqrt{2}} V^{\phantom{\ast}}_{ud}V^\ast_{us} 
\, {2\over 9}\, f_\pi &&\!\!\!\!\!\!\! \Biggl\{
   5\  g_{27} \, (M_K^2-M_\pi^2)\,\left[ 
 1+\Delta_L{\cal A}_2^{(27)} +\Delta_C{\cal A}_2^{(27)} \right]
  \Biggr.\nonumber\\
&& \!\!\!\! \Biggl.\mbox{}
  - 3\  e^2\ f^2_\pi \,  g_8 \,\left[
 g_{ew}\,\left( 1 +\Delta_L{\cal A}_{2}^{(ew)}\right)
  +\Delta_C{\cal A}_{2}^{(ew)} 
 \right] \Biggr\}\quad 
\eea 
for the $I=2$ amplitude. 
The electroweak penguin contributions have been also included. 
These  formulae contain
chiral loop corrections  $\Delta_L\cA_I^{(R)}$, 
coming from the lowest--order 
Lagrangian (\ref{eq:lg8_g27}) and its strong counterpart.
Loop corrections are always subleading in the $1/N_C$ expansion,
so that they do not enter the large--$N_C$ matching procedure
outlined in the introduction. One-loop corrections to $K\to\pi\pi$
have been extensively analyzed in ref.~\cite{PP:00b}, with the aim of
identifying and resum FSI effects. Those effects, subleading in $1/N_C$ but
numerically relevant, will be taken into account in sections~\ref{sec:loops}
and \ref{sec:FSI}.

At next--to--leading order in the chiral expansion, i.e. $\cO(G_F p^4)$ and
$\cO(G_Fe^2p^2)$, the complete Lagrangian which mediates non--leptonic weak
interactions with $\Delta S=1$ can be written as follows
\cite{KA91,BPP,EIMNP:00,CDG:99,CG:00,EKW:93,dR:89}:
\beq\label{eq:l4g8_g27}
\cL_4^{\Delta S=1} = -{G_F \over \sqrt{2}}  V_{ud}^{\phantom{*}} V_{us}^*
\, f^2\, \Bigg ( g_8\; \sum_i\; E_i\; O^8_i +
g_{27}\;\sum_i\; D_i\; O^{27}_i
   + g_8 \, e^2 f^2 \; \sum_i\; Z_i\; O^{EW}_i \Bigg ) \, .
\eeq
For the octet and 27--plet weak operators $O^8_i$ and $O_i^{27}$ the basis
constructed in ref. \cite{BPP} has been adopted\footnote{
For the octet operators one can use either  the basis of ref.~\cite{BPP} 
or the basis of ref.~\cite{EKW:93}. For completeness we provide
the transformation rules  between the two bases in appendix~\ref{APP_D}.}.
For the electroweak operators $O^{EW}_i$ we use the basis\footnote{
Our operators $O^{EW}_i$ are denoted with $Q_i$ in
ref.~\protect\cite{EIMNP:00} and their
coupling $G_8$ is related to our $g_8$ via the identity
$G_8 = -\left(G_F/\sqrt{2}\right) V_{ud}V_{us}^\ast\, g_8$.
}
of ref.~\cite{EIMNP:00}.
We refer to those references for the explicit form of the operators.

The $\cO(p^4)$ and $\cO(e^2p^2)$ tree--level contributions to the 
$K\to\pi\pi$ amplitudes are easily computed
with the Lagrangian (\ref{eq:l4g8_g27}) and its strong counterpart.
The complete expressions can also be obtained from refs.~\cite{BPP}
and \cite{EIMNP:00}:

\bea
\Delta_C\cA_0^{(8)} &\! =&\! \tilde{\Delta}_C
 +{2M_K^2\over f_\pi^2}\,\left ( E_{10}-2E_{13}+E_{15}\right) \nonumber\\
&&\hspace{-1.5cm}\mbox{}
+{2M_\pi^2\over f_\pi^2}\,\left (-2E_1-4E_2-2 E_3 +2E_{10}+E_{11}+4E_{13}\right)
 \, , \\ 
\Delta_C\cA_0^{(27)} &\! =&\!\tilde{\Delta}_C
+{M_K^2\over f_\pi^2}\,\left ( D_4-D_5-9D_6+4D_7\right)
\nonumber\\ &&\hspace{-1.5cm}\mbox{}
+{2M_\pi^2\over f_\pi^2}\,\left ( -6D_1-2D_2+2D_4+6D_6+D_7\right) \, , \\
\Delta_C\cA_{0}^{(ew)} &\! =&\!g_{ew}\,\tilde{\Delta}_C^{(ew)}+
\fr{2M_K^2}{ f_\pi^2} \left( Z_1+ 2 Z_2\right )
+\fr{M_\pi^2}{ f_\pi^2} \left(4 Z_1+ 2 Z_2 -Z_6\right )
\\
& &\hspace{-1.5cm}\mbox{}
-\fr{M_K^2-M_\pi^2}{6 f_\pi^2} \left(8 Z_3 - 24 Z_4 + 9 Z_5 + 6 Z_7 
- 3 Z_8 - 3 Z_9 - 2 Z_{10} + 2 Z_{11} + 2 Z_{12}\right)
 \, , \nn 
\label{eq:AMP_4} \\
\Delta_C\cA_{2}^{(27)} &\! =&\!\tilde{\Delta}_C
+{M_K^2\over f_\pi^2}\,\left( D_4-D_5+4D_7\right)
+{2M_\pi^2\over f_\pi^2}\,\left ( -2D_2+2D_4+D_7\right) \, ,
 \\
\Delta_C\cA_{2}^{(ew)} &\! =&\!g_{ew}\,\tilde{\Delta}_C^{(ew)}
+\fr{M_K^2}{ f_\pi^2}\,\left( 2 Z_1  + 4 Z_2 -  Z_6\right )
+\fr{2M_\pi^2}{ f_\pi^2}\,\left( 2Z_1  + Z_2 \right )
\nn \\ & & \hspace{-1.5cm}\mbox{}
+\fr{M_K^2-M_\pi^2}{ 3 f_\pi^2}\,\left( - 4 Z_3 + 12 Z_4- 3 Z_8 - 3 Z_9
- 2 Z_{10} + 2 Z_{11} + 2 Z_{12}   \right)
\, ,
\eea
where
\bea\label{eq:Deltas_def}
\tilde{\Delta}_C &=& -\fr{4 L_5}{f_\pi^2}\, \left(M_K^2+3 M_\pi^2\right)
-\fr{16 L_4}{f_\pi^2}\, \left(2 M_K^2+ M_\pi^2\right) \, ,
\nn\\
\tilde{\Delta}_C^{(ew)} &=& 
-\fr{4 L_5}{f_\pi^2}\, \left(M_K^2+5 M_\pi^2\right)
-\fr{24 L_4}{f_\pi^2}\, \left(2 M_K^2+ M_\pi^2\right) \, .
\eea

There are seven $(8_L,1_R)$ operators $O^8_i$
($i=1,2,3,10,11,13,15$), six $(27_L,1_R)$  operators $O^{27}_i$
($i=1,2,4,5,6,7$)
and twelve electroweak operators $O^{EW}_i$
($i= 1,\,\ldots\, ,12$) contributing to $K\to \pi\pi$
matrix elements \cite{BPP,EIMNP:00}.
The practical limitation of a systematic $\chi$PT evaluation of the
$K\to\pi\pi$ isospin amplitudes is in the fact that the counterterms
which appear at next-to-leading order are not fully known and their
determination would require the experimental knowledge of a large set 
of weak $\Delta S=1$ processes.

In addition, there are contributions involving the lowest--order
$\Delta S=1$\ Lagrangian (\ref{eq:lg8_g27}) combined with the $\cO(p^4)$
strong chiral operators with couplings $L_i$, introduced in
ref.~\cite{GL:85}.
In previous analyses \cite{KA91,BPP,CDG:99,CG:00}
these corrections, shown in eqs.~\eqn{eq:Deltas_def}, 
were factorized as global factors in front of the corresponding
amplitudes: \ 
$1+\tilde{\Delta}_C \,\dot=\, f^4/(f_\pi^3 f_K) \,\approx\, 0.65$,
$1+\tilde{\Delta}_C^{(ew)} \,\dot=\, f^6/(f_\pi^5 f_K) \,\approx\, 0.58$.
A factor $f^3/(f_\pi^2 f_K)$ arises from wave--function 
re\-nor\-ma\-li\-za\-tion,
while the remaining powers of $f/f_\pi$ are needed to rewrite
in terms of the physical pion decay constant
the explicit dependences of the tree--level amplitudes \eqn{eq:AMP_2}
on the chiral Lagrangian coupling $f$.
This procedure induces a sizeable suppression which is 
finally compensated by large and positive corrections from the
$\cO(p^4)$ weak counterterms.
We prefer to keep all $\cO(p^4)$ local contributions together and
perform a consistent large--$N_C$ calculation of their global size.

\section{Large--$N_C$ Matching}
\label{sec:matching}

In the large--$N_C$ limit the T--product of two colour--singlet
quark currents factorizes:
\bea\label{eq:factj}
\langle J \cdot J \rangle &=&
\langle J \rangle\;\langle J \rangle\; 
\left\{ 1 \, + \,\cO\left(1\over N_c \right)\right\}
\ .
\eea
In other words, colour exchanges between the two currents $J$ are
$1/N_C$ suppressed and in this limit the factorization of four--quark
operators is exact.
Since quark currents have well--known realizations in $\chi$PT 
\cite{GL:85,EC:95,PI:95},
the hadronization of the weak operators $Q_i$ can then
be done in a quite straightforward way.
Thus, at large--$N_C$ the matching between the short--distance
Lagrangian \eqn{eq:lag} and its long--distance $\chi$PT realization
can be explicitly performed.

The chiral couplings of the lowest--order
Lagrangian~(\ref{eq:lg8_g27}) have the following large--$N_C$ values:
\bea\label{eq:c2}
g_8^\infty&=& -{2\over 5}\,C_1(\mu)+{3\over 5}\,C_2(\mu)+C_4(\mu)
- 16\, L_5\,\left( {\langle\bar q q \rangle^{(2)}(\mu) \over f^3}\right)^2
\,C_6(\mu)\, ,
\nonumber\\
g_{27}^\infty&=&{3\over 5}\,[C_1(\mu)+C_2(\mu)]\, ,
\\
(g_8\, e^2 g_{ew})^\infty&=& -3\,
\left( {\langle\bar q q \rangle^{(2)}(\mu) \over f^3}\right)^2
\, C_8(\mu)\, .
\nonumber\eea

Together with the $\cO(p^2)$ amplitudes in eqs.~\eqn{eq:AMP_2}, these
results are equivalent to the standard large--$N_C$ evaluation of the
usual bag parameters $B_i$.
In particular, for $\eps'/\eps$, where only the imaginary part of the
$g_i$ couplings matter [i.e. Im($C_i$)], eqs.~\eqn{eq:c2} 
amount to $B_8^{(3/2)}\approx B_6^{(1/2)}=1$. Therefore, up to minor
variations on some input parameters, the corresponding $\eps'/\eps$
prediction, obtained at lowest order in both the $1/N_C$ and
$\chi$PT expansions, reproduces the published results of the Munich
\cite{munich} and Rome \cite{rome} groups.

The large--$N_C$ limit has been only applied to the matching between
the 3--flavour quark theory and $\chi$PT,
as indicated in figure~\ref{fig:eff_th}.
The evolution from the electroweak
scale down to $\mu < m_c$ has to be done without any unnecessary expansion
in powers of $1/N_C$; otherwise, one would miss large corrections
of the form ${1\over N_C} \ln{(M/m)}$, with $M\gg m$ two widely
separated scales \cite{BBG87}.
Thus, the Wilson coefficients contain the full $\mu$ dependence.

The operators $Q_i$ ($i\not=6,8$) factorize into products of 
left-- and right--handed vector currents, 
which are renormalization--invariant quantities.
The matrix element of each single current
represents a physical observable which can be directly measured;
its $\chi$PT realization just provides a low--energy expansion in
powers of masses and momenta.
Thus, the large--$N_C$ factorization of these operators
does not generate any scale dependence. Since the anomalous
dimensions of $Q_i$ ($i\not=6,8$) vanish when $N_C\to\infty$ \cite{BBG87},
a very important ingredient is lost in this limit \cite{PI:89}.
To achieve a reliable expansion in powers of $1/N_C$,
one needs to go to the next order where this physics is captured
\cite{PI:89,PR:91}. This is the reason why the study of the $\Delta I=1/2$
rule has proved to be so difficult. Fortunately, these operators
are numerically suppressed in the $\eps'/\eps$ prediction.

The only anomalous dimensions which survive when $N_C\to\infty$
are the ones corresponding to $Q_6$ and $Q_8$ \cite{BBG87,BG:87}.
One can then expect that the matrix elements of these two operators
are well approximated by this limit\footnote{
Some insight on these matrix elements can be obtained from
the two-point functions $\Psi_{ii}(q^2)\equiv i \int d^4x\, e^{iqx}\, 
\langle T(Q_i(x) Q_i(0)^\dagger)\rangle$,
since their absorptive parts correspond to an inclusive sum of
hadronic matrix elements squared. The known $\cO(\alpha_s)$ results
\protect\cite{PI:89,PR:91,JP:94} show that the large--$N_C$ limit provides
an excellent approximation to $\Psi_{66}$, but an incorrect description
of $\Psi_{22}$.
} \cite{PI:89,PR:91,JP:94}.
These operators  factorize into colour--singlet
scalar and pseudoscalar currents, which are $\mu$ dependent.
Since the products $m_q \;\bar{q}(1,\g_5) q$,
are physical observables, the scalar and pseudoscalar currents
depend on $\mu$ like the inverse of a quark mass.
Conversely,  the Wilson coefficients of the operators
$Q_6$ and $Q_8$ scale with $\mu$ like the square of
a quark mass in the large--$N_C$ limit.

The $\chi$PT evaluation of the scalar and pseudoscalar currents provides,
of course, the right $\mu$ dependence, since only physical observables
can be realized in the low--energy theory. What one actually
finds is the chiral realization of the renormalization--invariant
products $m_q \;\bar{q}(1,\g_5) q$.
This generates the factors [$m_q \equiv m_u = m_d$]
\bea\label{eq:B0_comp}
{\langle\bar q q \rangle^{(2)}(\mu) \over f^3}
 \!\!\! &\equiv&\!\!\!  -{B_0\over f} \, = \, 
 -{B_0\over f_\pi}\; {f_\pi \over f} 
\nn\\
& = &\!\!\!
- {M_\pi^2\over 2\, m_q(\mu)\, f_\pi}\; 
 \Biggl[ 1 + {4 L_5\over f_\pi^2}\, M_\pi^2 
 + 4 \,\fr{2 M_K^2+M_\pi^2}{f_\pi^2}(3 L_4-4 L_6)
 -8 \,\fr{M_\pi^2}{f_\pi^2}\, (2 L_8-L_5)
\Biggr.
\nn \\
& & \Biggl.\qquad\qquad\quad\;\mbox{}
-3 \,\nu_\pi - \nu_K - \fr{1}{3} \,\nu_\eta\Biggr] \nn
\\ & = &\!\!\!
 - {M_K^2\over (m_s + m_q)(\mu)\, f_\pi}\;
 \Biggl[ 1 + {4 L_5\over f_\pi^2}\, M_\pi^2 
  + 4 \,\fr{2 M_K^2+M_\pi^2}{f_\pi^2}(3 L_4-4 L_6)
\Biggr.
\nn \\ & & \Biggl.\qquad\qquad\qquad\qquad\mbox{}
 -8 \,\fr{M_K^2}{f_\pi^2}\, (2 L_8-L_5)
-2 \,\nu_\pi - \nu_K - \fr{2}{3} \,\nu_\eta\Biggr] \, , 
\eea
in eqs.~\eqn{eq:c2}, which exactly cancel the $\mu$ dependence of
$C_{6,8}(\mu)$ at large $N_C$ 
\cite{BBG87,BG:87,dR:89,PI:89,PR:91,JP:94}.
It remains a dependence at next-to-leading order.
The parameter $B_0$ is a low--energy coupling of the $\cO(p^2)$
strong chiral Lagrangian, which accounts for the vacuum quark
condensate at lowest order in the momentum expansion.
The one-loop corrections $\nu_P$\ ($P=\pi,\, K,\,\eta$), 
defined in appendix~\ref{APP_B},
are identically zero in the limit $N_C\to\infty$.

While the real part of $g_8$ gets its main contribution
from $C_2$, Im($g_8$) and Im($g_8\,g_{ew}$)
are governed  by $C_6$ and $C_8$, respectively.
Thus, the analyses of the CP--conserving and CP--violating amplitudes
are very different. There are large $1/N_C$ corrections to Re($g_i$)
\cite{PI:89,PR:91,JP:94}, which are needed to understand
the observed enhancement of the $(8_L,1_R)$ coupling.
However, the large--$N_C$ limit can be expected to give a
good estimate of Im($g_i$).

Contrary to the other $Q_i$ operators, the leading--order contribution
of $Q_6$ involves the coupling $L_5$ of the 
${\cal O}(p^4)$  strong chiral Lagrangian. The large--$N_C$ value
of this chiral coupling can be estimated from the ratio of the kaon and
pion decay constants:
\beq
L_5^\infty = {f_\pi^2\over 4 \, \left(M_K^2-M_\pi^2\right)}\,
\left( {f_K\over f_\pi} - 1 \right) = 2.1\cdot 10^{-3} \, .
\eeq
The $Q_6$ contribution dominates the numerical value of Im($g_8^\infty$).
In the large--$N_C$ limit, the combined effect of all other operators 
only amounts to a 5\% correction.

The $\cO(p^4)$ corrections introduce dependences on three additional
strong chiral couplings. At large $N_C$, 
\beq
L_{4}^\infty \: = \: L_{6}^\infty \: = \: 0\, .
\eeq
To determine $L_8$,
we impose the stronger requirement of lowest--meson dominance
\cite{EGPR:89,PPR:98} and assume that the scalar form factors
vanish at infinite momentum transfer. This
implies the relation\cite{JOP:00}
\beq
(2 L_8 - L_5)^\infty \: = \: 0\, ,
\eeq
which is well satisfied by the phenomenological 
determinations of those constants \cite{GL:85,ABT:00}.

The operators $Q_3$ and $Q_5$ start to contribute at $\cO(p^4)$, while
the electroweak penguin operators $Q_7,\, Q_9$ and $Q_{10}$ give their
first contributions at $\cO(e^2p^2)$.
The large--$N_C$ matching at the next-to-leading chiral order fixes the
couplings $E_i$, $D_i$ and $Z_i$ of the
long--distance chiral Lagrangian (\ref{eq:l4g8_g27}).
We  only quote the values of those couplings contributing to
$K\to\pi\pi$ amplitudes.

For the $\cO(p^4)$ couplings, one gets:

\bea
\label{eq:cw4}
(g_8 E_{1})^\infty &=& -48\ \ds X_{19} \CQ^2 C_6(\mu) \, ,\nn \\
(g_8 E_{2})^\infty &=& -32\ \ds X_{20} \CQ^2 C_6(\mu) \, , \nn \\
(g_8 E_{3})^\infty &=& -16\ \ds X_{31} \CQ^2 C_6(\mu) \, , \nn \\
(g_8 E_{10})^\infty&=& 2\ L_5\, \left[-\fr{2}{5}C_1(\mu)+\fr{3}{5} C_2(\mu)+
              C_4(\mu) \right]
\nn \\ & & 
-8\ \ds (2 X_{14}+2 X_{15}+X_{38}) \CQ^2 
C_6(\mu)\, ,
\nonumber\\
(g_8 E_{11})^\infty&=& 4\ L_5\,\left[-\fr{2}{5}C_1(\mu)+\fr{3}{5} C_2(\mu)+
              C_4(\mu) \right]
\nn \\ & & 
-16\ \ds (X_{15}+2 X_{17}-X_{38})  \CQ^2 C_6(\mu)\,  ,
\nonumber\\
(g_8 E_{13})^\infty &=& 8\  \ds (X_{15}-4 X_{16}) \CQ^2 C_6(\mu) \, , \nn \\
\nonumber\\
(g_8 E_{15})^\infty &=& 8\ \ds (-X_{34}+X_{38}) \CQ^2 C_6(\mu) \, , \nn \\
(g_{27} D_4)^\infty&=& 4\  L_5\, g_{27}^\infty \, .
\eea
All other $(27_L,1_R)$ couplings contributing to $K\to \pi \pi$
($D_1$, $D_2$, $D_5$, $D_6$ and $D_7$) are zero at large--$N_C$.

The $\cO(p^4)$ contributions from the operator $Q_6$ have been computed using
the $\cO(p^6)$ Lagrangian of ref.~\cite{BCE:99}; the couplings $X_i$ refer
to the list of $\cO(p^6)$ $SU(3)$ operators given in that reference.
These couplings however are unknown, so in practice
the $Q_6$ contribution is missing in eqs.~\eqn{eq:cw4}.
The remaining terms are in agreement with the results obtained in
ref.~\cite{EKW:93}.

The non-zero $\cO(e^2p^2)$  couplings relevant for $K\to\pi\pi$ are:
\bea
 (g_8 e^2 Z_1)^\infty&=& -24\,
\left( {\langle\bar q q \rangle^{(2)}(\mu) \over f^3}\right)^2
\,L_8\,C_8(\mu)\, ,
\nonumber\\
(g_8 e^2 Z_5)^\infty&=& C_{10}(\mu) \, ,
\nonumber\\
 (g_8 e^2 Z_6)^\infty&=& -12\,
\left( {\langle\bar q q \rangle^{(2)}(\mu) \over f^3}\right)^2\,
L_5\,C_8(\mu)\, , 
\nonumber\\
(g_8 e^2 Z_8)^\infty&=& {3\over 2}\,
\left[ C_9(\mu)+C_{10}(\mu)\right]\, ,
\nonumber\\
(g_8 e^2 Z_9)^\infty&=& -{3\over 2}\, C_7(\mu)\, .  
\eea

\section{Isospin Amplitudes at Leading Order in $1/N_C$}
\label{sec:N_C_results}

Combining the results of the previous sections, one gets
the predicted $K\to\pi\pi$ amplitudes at leading order in $1/N_C$.
The different contributions to the
isospin amplitudes take the following form:
\bea\label{eq:NC_results}
\lefteqn{
g_8^\infty\,\left[ 1 + \Delta_C\cA_0^{(8)} \right]^\infty\, =} &&
\nonumber\\ &&
\left\{ 
 -{2\over 5}\,C_1(\mu)+{3\over 5}\,C_2(\mu)+C_4(\mu)
- 16\, L_5\, C_6(\mu)\,
\left[ {M_K^2 \over (m_s + m_q)(\mu)\, f_\pi}\right]^2\right\}\,
f_0^{K\pi}(M_\pi^2)   \, ,\qquad\quad
\\  &&\nonumber\\ \nonumber\\
\lefteqn{
g_{27}^\infty\,\left[ 1 + \Delta_C\cA_0^{(27)}\right]^\infty \ = \
g_{27}^\infty\,\left[ 1 + \Delta_C\cA_2^{(27)}\right]^\infty \, = \,
{3\over 5}\,[C_1(\mu)+C_2(\mu)]\;  f_0^{K\pi}(M_\pi^2) \, , }&&
\\ &&\nonumber\\ &&\nonumber\\
\lefteqn{
e^2\, g_8^\infty\, \left[ g_{ew} + \Delta_C\cA_0^{(ew)}\right]^\infty 
\, =\,
-3\, C_8(\mu)\,\left[ {M_K^2 \over (m_s + m_q)(\mu)\, f_\pi}\right]^2
\left[1 + {4 L_5\over f_\pi^2} M_K^2\right]
} &&
\nonumber\\ &&\hspace{4.3cm}\mbox{}
- {3\over 4} \left[ C_7- C_9 + C_{10}\right]\!(\mu) \:
{M_K^2-M_\pi^2\over f_\pi^2}\; f_0^{K\pi}(M_\pi^2)\, ,
\\ &&\nonumber\\ \nonumber\\
\lefteqn{
e^2\, g_8^\infty\, \left[ g_{ew} + \Delta_C\cA_2^{(ew)}\right]^\infty
 \, = \,
-3\, C_8(\mu)\,\left[ {M_K^2 \over (m_s + m_q)(\mu)\, f_\pi}\right]^2
\left[1 + {4 L_5\over f_\pi^2} M_\pi^2\right]
}&&\hfill\mbox{}
\nonumber\\ &&\hspace{4.3cm}\mbox{}
+ {3\over 2} \left[ C_7 - C_9 - C_{10}\right]\!(\mu)\:
{M_K^2-M_\pi^2\over f_\pi^2}\; f_0^{K\pi}(M_\pi^2) \, .
\eea

For the operators $Q_i$ ($i\not= 6,8$), which are products of 
colour--singlet vector and axial--vector currents, these are exact 
large--$N_C$ results to all orders in the chiral expansion, as can
be easily seen factorizing the operators at the quark level.
The $\chi$PT framework discussed before reproduces these results in
a perturbative way, through the momentum expansion of the $K\pi$ scalar
form factor at $\cO(p^4)$:
\beq
f_0^{K\pi}(M_\pi^2)\:\equiv\: f_+^{K\pi}(M_\pi^2) +
{M_\pi^2\over M_K^2 - M_\pi^2}\: f_-^{K\pi}(M_\pi^2)
\: = \: 1 + {4 L_5\over f_\pi^2}\, M_\pi^2 + \cdots
\eeq
The form factors $f_\pm^{K\pi}(t)$ are defined through the
matrix element of the vector current,
\beq
\langle\pi| \bar s \gamma^\mu q |K\rangle \, = \,
C_{K\pi}\;\left\{ \left( P_K + k_\pi\right)^\mu\; f_+^{K\pi}(t)
+ \left( P_K - k_\pi\right)^\mu\; f_-^{K\pi}(t) \right\}
\qquad\quad (q=u,\, d),
\eeq
where $t\equiv \left( P_K - k_\pi\right)^2$, 
$C_{K^0\pi^0}= -C_{K^+\pi^0}= 1/\sqrt{2}$ and 
$C_{K^0\pi^-} =C_{K^+\pi^+}= -1$.

The wave--function renormalization corrections $\tilde{\Delta}_C$
[eq.~\eqn{eq:Deltas_def}]
have been cancelled by weak $\cO(p^4)$ contributions, as it should
since we are dealing with conserved currents.
Once the $\cO(p^2)$ results are written in terms of the physical
pion decay constant $f_\pi$, 
higher--order chiral contributions only introduce the small
correction factor
$f_0^{K\pi}(M_\pi^2)\approx 1.02$.

The hadronic matrix elements of the operators $Q_6$ and $Q_8$
factorize into products of scalar and pseudoscalar currents,
which cannot be directly measured.
The $\chi$PT predictions are then needed to determine those
hadronic currents.
The electroweak penguin matrix elements are known to $\cO(p^4)$.
Again, one observes that the contributions from local weak terms
($Z_1$ and $Z_6$)
cancel the negative contribution from $\tilde{\Delta}_C^{(ew)}$
and reverse the sign of the $\cO(p^4)$ correction.
The contribution of the penguin operator $Q_6$ is only known at
$\cO(p^2)$. For $Q_6$ we cannot just include the $\tilde{\Delta}_C$
correction, because the corresponding weak $\cO(p^4)$ counterterms
are unknown and large cancellations can be expected.
In eq.~\eqn{eq:NC_results} we have taken a global correction
factor $f_0^{K\pi}(M_\pi^2)$ for the octet amplitude.
This is a reasonable assumption\footnote{
In fact,
the factor $f_0^{K\pi}(M_\pi^2)$ already appears in the
lowest--order $Q_6$ contribution to $g_8$, through the
$\cO(p^4)$ correction in eq.~\eqn{eq:B0_comp}.
},
since nearly all known pieces have
this common correction.
Only $\langle Q_8\rangle_0$ gets a different (and larger) correction.

\begin{table}[tbh]\centering
\caption{Numerical values of the weak chiral couplings
in the large--$N_C$ limit.}
\vspace{0.3cm}
\label{tab:N_C_predictions}
\begin{tabular}{|c|c|}
\hline
$g_8^\infty\,\left[ 1 + \Delta_C\cA_0^{(8)} \right]^\infty$ &
$\left( 1.3\pm 0.2\, \pm 0.1\right) + \tau\:
\left( 1.12\pm 0.08\, {}^{+0.49}_{-0.30}\right)$
\\
$g_{27}^\infty\,\left[ 1 + \Delta_C\cA_0^{(27)}\right]^\infty$ &
$\left( 0.47\pm 0.01\pm 0.0\right)$
\\
$e^2\, g_8^\infty\, 
\left[ g_{ew} + \Delta_C\cA_0^{(ew)}\right]^\infty$ &
$-\left( 0.085\pm 0.085\, {}^{+0.035}_{-0.023}\right) - \tau\:
\left( 2.33\pm 0.07\, {}^{+0.83}_{-0.52}\right)$
\\
$e^2\, g_8^\infty\, 
\left[ g_{ew} + \Delta_C\cA_2^{(ew)}\right]^\infty$ &
$-\left( 0.07\pm 0.07\, {}^{+0.03}_{-0.02}\right) - \tau\:
\left( 1.34\pm 0.03\, {}^{+0.68}_{-0.42}\right)$
\\ \hline
\end{tabular}
\end{table}

The scalar and pseudoscalar currents introduce a quadratic dependence
on quark masses in the contributions from the operators
$Q_6$ and $Q_8$.
At present, the most reliable determinations of the light quark masses
give $m_s(1~\mbox{GeV}) = (150\pm 25)$ MeV 
\cite{ms2000,PP:99,ALEPH:99,KM:00,JA:98,LatMass}
and $(m_u + m_d)(1~\mbox{GeV}) = (12.8\pm 2.5)$ MeV \cite{BPR:98},
at the scale $\mu=1$ GeV. We then take:
\beq
(m_s + m_q)(1~\mbox{GeV}) = (156\pm 25)\;\mbox{MeV}\, .
\eeq

Table~\ref{tab:N_C_predictions} shows the resulting numerical 
predictions for the weak chiral couplings. 
The central values have been obtained at $\mu=1$ GeV.
The first errors indicate the sensitivity to changes of the
short--distance renormalization scale in the
range $M_\rho<\mu<m_c$ and to the choice of $\gamma_5$ scheme 
in the next-to-leading order calculation of the Wilson coefficients.
The second uncertainties correspond to the input values
of the quark masses.

For historical reasons, the values of the
short--distance Wilson coefficients are usually given in terms of
$\Lambda_{\mathrm QCD}$ (in the three or four flavour theory).
Nowadays, that $\as$ is experimentally known with rather good accuracy,
it is unnecessary to introduce this additional auxiliary parameter
which only complicates the final expressions.
Since the most important $\as$ corrections appear at the lowest scale
$\mu\sim \cO (1 \ {\rm GeV})$,
we have fixed the strong coupling at the $\tau$ mass,
where it is known~\cite{pich1} with about a few percent level of accuracy:
\beq \label{eq:atau}
\a_s(M_\tau)=0.345\pm 0.020 \ .
\label{eq:amtau}
\eeq
The high--energy matching scale is chosen to be intermediate between the
$W$--boson and the top quark mass scale.  We have  performed the matching
directly at the $Z$--boson mass scale where $\a_s$ is best known~\cite{PDG:00},
\beq \label{eq:aZ}
\a_s(M_Z)=0.119\pm 0.002 \ .
\label{eq:amz}
\eeq
The measured values \eqn{eq:atau} and \eqn{eq:aZ} are in perfect agreement,
if one performs \cite{RPS:98} a four--loop evolution
of $\a_s$ between $M_Z$ and $M_\tau$, with the appropriate matching
conditions at the different thresholds \cite{ChKS:97}.
The values of $\a_s$ at the other needed scales can be
deduced from \eqn{eq:atau}.
The numerical uncertainties associated with the present error on $\as$
have been included in our results, but they are negligible in
comparison with the uncertainties from other sources.

The dominance of $Q_6$ and $Q_8$ in the CP--odd amplitudes (the ones
proportional to the CKM factor $\tau$) is apparent in
table~\ref{tab:N_C_predictions}, where those pieces show a very
strong dependence on quark masses (second error bars).
In comparison, the short--distance uncertainties are much smaller.
The opposite behaviour is observed in the CP--conserving couplings
Re$(g_{8})$ and Re$(g_{27})$,
which are dominated by $Q_1$ and $Q_2$.
The 27--plet coupling, which does not get any penguin contribution,
satisfies Im$(g_{27}) =0$ for all practical purposes.

Taking $\Omega_{IB}^{\pi^0\eta} = 0.16$, 
Im$(V_{ts}^* V_{td}^{\phantom{*}}) = 1.2\cdot 10^{-4}$
and the central values in table~\ref{tab:N_C_predictions}
for the CP--odd amplitudes,
one gets the large--$N_C$ prediction
Re$(\varepsilon'/\varepsilon) \approx 0.8\cdot 10^{-3}$.
Although numerically suppressed, the operators $Q_1$, $Q_2$
and $Q_4$, which are not well approximated by the large--$N_C$
limit, provide also small corrections to Im$(A_0)$. In
refs.~\cite{munich,buras1}
the measured CP--conserving rates are used to estimate those
contributions. This amounts to multiply the corrections
from these operators by a factor
$\xi_0 \approx 4.9$, to compensate for the
underestimated coupling Re$(g_8)$.
Adopting this prescription, one gets
Re$(\varepsilon'/\varepsilon) \approx 0.5\cdot 10^{-3}$,
in agreement with the findings of refs.~\cite{munich,rome}.

\section{Chiral Loop Corrections}
\label{sec:loops}

The previous tree--level amplitudes do not contain any strong phases
$\delta^I_0$.
Those phases originate in the final rescattering of the two pions and,
therefore, are generated by chiral loops which are of higher order in
the $1/N_C$ expansion.
Analyticity and unitarity require the presence of a corresponding
dispersive FSI effect in the moduli of the isospin amplitudes.
Since the strong phases are quite large, specially in the isospin--zero
case, one should expect large higher--order unitarity corrections.
Intuitively, the behaviour of the $I=0$ and $I=2$ S--wave phase shifts
as a function of the total invariant mass of the two pions
suggests a large enhancement of the $I=0$
amplitude and a small suppression of the $I=2$ amplitude.

The size of the FSI effect can be estimated at one loop in $\chi$PT.
The dominant one-loop correction to the octet amplitude
comes indeed from the
elastic soft rescattering of the two pions in the final state.
The existing one-loop analyses~\cite{KA91,BPP} show
that pion loop diagrams provide an important enhancement
of the $A_0$ amplitude by about $40\%$, implying a
sizeable reduction of the phenomenologically fitted value of
$|g_8|$ in eq.~(\ref{eq:g8_g27}).

The complete formulae for the one-loop corrections $\Delta_L{\cal A}_I^{(R)}$
are compiled in the appendices.
The usual one-loop function $B(M_1^2,M_2^2, p^2)$
is defined in appendix \ref{APP_B}, while appendix~\ref{sec:onshell}
contains explicit results for the different isospin amplitudes.
The contributions proportional to $B(M_P^2,M_P^2,M_K^2)$, with 
$P=\pi ,\, K,\, \eta$, arise from 
intermediate $\pi\pi$, $K\bar{K}$ and $\eta\eta$ states. 
At $s\equiv \left(p_{\pi_1}+p_{\pi_2}\right)^2 = M_K^2$, the only possible
absorptive contribution comes from the elastic $\pi\pi$ rescattering:
\bea
\Delta_L\cA_0^{(R)} & = & -{1\over 2} \left(2 M_K^2-M_\pi^2\right)\;
 B(M_\pi^2,M_\pi^2,M_K^2)\; + \;\ldots
\\
\Delta_L\cA_2^{(R)} & = & \phantom{-}{1\over 2} \left(M_K^2-2 M_\pi^2\right)
\; B(M_\pi^2,M_\pi^2,M_K^2)\; + \;\ldots
\eea
where
\beq
B(M_\pi^2,M_\pi^2,M_K^2) \; =\; {1\over (4\pi f_\pi)^2} \left\{
\sigma_\pi\,\left[
\ln{\left( {1+\sigma_\pi\over 1-\sigma_\pi}\right)}
 - i \,\pi \right]
- \ln{\left({\nu^2\over M_\pi^2}\right)}- 1 \right\} \, ,
\eeq
with $\nu$ the chiral loop renormalization scale and
\beq
\sigma_\pi\;\equiv\;\sqrt{1-{4 M_\pi^2\over M_K^2}} \, .
\eeq

Thus, all isoscalar amplitudes get the same absorptive contribution,
as it should, since they have identical strong phase shifts.
The same is true for the two amplitudes with $I=2$.
The one-loop absorptive contributions reproduce
the leading $\chi$PT values of the strong rescattering phases 
 $\delta^I_0$, with $I=0,2$:
\beq\label{eq:CHPT_delta} 
\tan\delta_0^{0;2}(M_K^2) = {1\over 32\pi f_\pi^2} \; \sigma_\pi \;\left ( 
  2M_K^2-M_\pi^2\, ;\,  2M_\pi^2-M_K^2\right )
\, . 
\eeq 
The numerical values of $\delta_0^{0;2}(M_K^2)$ predicted by $\chi$PT 
at leading order, $\delta_0^{0}(M_K^2) = 25^\circ$ and 
$\delta_0^{2}(M_K^2) = -12^\circ$,
are significantly lower than their experimental values, implying 
that higher--order rescattering contributions are numerically relevant.
The phase-shift difference, 
$\delta_0^{0}-\delta_0^{2}= 37^\circ$, is slightly less sensitive to 
higher--order chiral corrections \cite{GM:91}.

The $2\pi$ intermediate state induces a large one-loop
correction to the $I=0$ amplitudes. At $\nu=M_\rho$, the
$2\pi$ contribution to the isoscalar amplitudes is
$\Delta_L\cA_0^{(R)}|_{\pi\pi} = 0.43 + 0.46\, i$, while
$\Delta_L\cA_2^{(R)}|_{\pi\pi} = -(0.19 + 0.20\, i)$; i.e.
the expected enhancement (suppression) of the $I=0$ ($I=2$)
amplitudes.
The contributions from other one-loop diagrams, not related to
FSI, are different for the different amplitudes
${\cal A}_I^{(R)}$.

Let us write our isospin amplitudes in the form
\beq
{\cal A}_I^{(R)} \; = \; {\cal A}_I^{(R)\infty} \;\times\;\:
{\cal C}_I^{(R)} \, ,
\eeq
where ${\cal A}_I^{(R)\infty}$ are the large--$N_C$ results
obtained in the previous section.
The correction factors ${\cal C}_I^{(R)}$ contain
the chiral loop contributions we are interested in.
At the one-loop level, they take the following
numerical values:
\bea
{\cal C}_0^{(8)}&\approx & 1 +
\Delta_L{\cal A}_0^{(8)} \;\;\: =\; 1.27 \pm 0.05 +  0.46 \ i \, ,\nn  \\
{\cal C}_0^{(27)}&\approx & 1 +
\Delta_L{\cal A}_0^{(27)} \,\; =\; 2.0 \pm 0.7 + 0.46 \ i \, ,\nn  \\
{\cal C}_0^{(ew)}&\approx & 1 +
\Delta_L{\cal A}_0^{(ew)} \; =\; 1.27  \pm 0.05 + 0.46 \ i \, ,\nn  \\
{\cal C}_2^{(27)}&\approx & 1 + 
\Delta_L{\cal A}_2^{(27)} \,\; =\; 0.96 \pm 0.05-0.20 \ i\, ,\nn  \\
{\cal C}_2^{(ew)}&\approx & 1 +
\Delta_L{\cal A}_2^{(ew)} \; =\; 0.50  \pm 0.24-0.20 \ i \, .
\label{eq:onel}
\eea
The central values have been evaluated at the chiral renormalization
scale $\nu = M_\rho$. To estimate the corresponding uncertainties
we have allowed the scale $\nu$ to change between 0.6 and 1~GeV.
The scale dependence is only present in the dispersive contributions
and should cancel with the corresponding $\nu$
dependence of the local counterterms. However, this dependence is
next-to-leading in $1/N_C$ and, therefore, is not included in our
large--$N_C$ estimate of the $\cO(p^4)$ and $\cO(e^2 p^2)$
chiral couplings.
The $\nu$ dependence of the chiral loops would be
cancelled by the unknown $1/N_C$--suppressed corrections
$\Delta_C{\cal A}_I^{(R)}(\nu)-\Delta_C{\cal A}_I^{(R)\infty}$,
that we are neglecting in the factors ${\cal C}_I^{(R)}$.
The numerical sensitivity of our results to the scale $\nu$
gives then a good estimate of those missing contributions.

The absorptive contribution induces a large one-loop
correction to the $I=0$ amplitudes. The
dispersive correction to $\Delta_L{\cal A}_0^{(27)}$
is even larger, but it has a smaller phenomenological impact
because the isoscalar $K\to\pi\pi$ amplitude is dominated by its octet
component; this 27--plet correction has a strong dependence
on $\nu$ and, therefore, a rather large uncertainty. 
Although the one-loop correction to the $I=2$ \ $(27_L,1_R)$
amplitude is rather moderate, the electroweak $I=2$ amplitude
gets a large dispersive correction with negative sign. This
induces a corresponding suppression of 
$|{\cal A}_2^{(ew)}|$ by about 46\%.

The numerical corrections to the 27--plet amplitudes do not have 
much phenomeno\-lo\-gi\-cal interest for CP--violating observables, 
because ${\rm Im}(g_{27}) = 0$.
Remember that the CP--conserving amplitudes ${\rm Re}(A_I)$
are set to their experimentally determined values.
What is relevant for the $\varepsilon'/\varepsilon$
prediction is the 35\% enhancement of the isoscalar octet
amplitude \ Im[$A_0^{(8)}$] \ and the 46\% reduction of \
Im[$A_2^{(ew)}$]. Just looking to the simplified formula
(\ref{EPSNUM}), one realizes immediately the obvious impact of
these one-loop chiral corrections, which destroy the
accidental lowest--order cancellation between the $I=0$
and $I=2$ contributions, generating a sizeable enhancement
of $\varepsilon'/\varepsilon$.

A complete $\cO(p^4)$ calculation \cite{EMNP:00,EIMNP:00}
of the isospin--breaking parameter $\Omega_{IB}$ is  not yet
available. The value 0.16 quoted in eq.~(\ref{eq:omIB}) 
only accounts for the contribution from $\pi^0$--$\eta$ 
mixing \cite{EMNP:00}
and should be corrected by the effect of chiral loops. 
Since $|{\cal C}_2^{(27)}| \approx 0.98\pm 0.05$,
one does not expect any large correction of \ ${\rm Im}(A_2)_{IB}$,
while we know that
Im[$A_0^{(8)}$] gets enhanced by a factor 1.35.
Taking this into account, one gets the corrected value
\beq
\Omega_{IB} \;\approx\;\Omega_{IB}^{\pi^0\eta}\;
\left|{{\cal C}_2^{(27)}\over {\cal C}_0^{(8)}}\right|
\; = \; 0.12 \pm 0.05 \, ,
\eeq
where the quoted error is an educated theoretical guess.
This value agrees with the result 
$\Omega_{IB} = 0.08 \pm 0.05\pm 0.01$,
obtained in ref.~\cite{MW:00} by using three different
models \cite{trieste,EKW:93,PR:91,EGPR:89,EPR:91,GV:99}
to estimate the relevant $\cO(p^4)$ chiral couplings.

The one-loop corrections increase the large--$N_C$ estimate
from $\varepsilon'/\varepsilon \,\approx\, 0.8\cdot 10^{-3}$
to\footnote{This number is obtained taking the experimental
value for $\eps$ and 
Im$\left( V_{ts}^* V_{td}^{\phantom{*}}\right)=1.2\cdot 10^{-4}$.
Using instead the theoretical prediction for $\eps$, one would get
 $\varepsilon'/\varepsilon \,\approx\, 2.2\cdot 10^{-3}$.
See section 8 for more details on this second kind of numerical analysis.}
$\varepsilon'/\varepsilon \,\approx\, 1.8\cdot 10^{-3}$.
The contributions to Im$(A_0)$ 
from the operators $Q_{1,2,4}$ can be
corrected phenomenologically, as advocated in ref.~\cite{buras1};
this requires now a smaller factor $\xi_0\approx 3.5$,
which results in\footnote{
Using the theoretical value of $\eps$, one finds
$\varepsilon'/\varepsilon \,\approx\, 1.8\cdot 10^{-3}$.}
$\varepsilon'/\varepsilon \,\approx\, 1.5\cdot 10^{-3}$.

\section{Final State Interactions at Higher Orders}
\label{sec:FSI}

Given the large size of the one-loop contributions, one should
worry about higher--order chiral corrections.
The fact that the one-loop calculation still underestimates
the observed $\delta_0^0$ phase shift indicates that a further enhancement
could be expected at higher orders.

The large one-loop FSI correction to the isoscalar amplitudes 
is generated by large infrared
chiral logarithms involving the light pion mass \cite{PP:00b}.
These logarithms are universal, i.e. their contribution depends exclusively 
on the quantum numbers of the two pions in the final state \cite{PP:00b}. 
As a result, they give the same correction to all isoscalar amplitudes.
Identical logarithmic contributions appear in the scalar pion form factor
\cite{GL:85}, where they completely dominate the $\cO(p^4)$  $\chi$PT
correction.

Using analyticity and unitarity constraints \cite{GP:97}, these
logarithms can be exponentiated to all orders in the chiral
expansion \cite{PP:00,PP:00b}.
The result can be written as:
\beq\label{eq:OMNES_WA}
{\cal C}_I^{(R)} \;\equiv\; {\cal C}_I^{(R)}(M_K^2)
\; =\;\Omega_I(M_K^2,s_0) \; {\cal C}_I^{(R)}(s_0)\, .
\eeq
The Omn\`es \cite{GP:97,OMNES,TR:88} exponential\footnote{
Equivalent expressions with an arbitrary number of subtractions for the
dispersive integral can be written \protect\cite{PP:00b}.
}
%
\beq\label{eq:omega}
\Omega_I(s,s_0) \;\equiv\;
e^{i\delta^I_0(s)}\; \Re_I(s,s_0) \; =\;
 \exp{\left\{ {(s-s_0)\over\pi}\int
{dz\over (z-s_0)} {\delta^I_0(z)\over (z-s-i\epsilon)}\right\}}
\eeq
provides an evolution of ${\cal C}_I^{(R)}(s)$ from an arbitrary
low--energy point $s_0$ to 
$s\equiv\left(p_{\pi_1}+p_{\pi_2}\right)^2=M_K^2$.
The physical amplitudes are of course independent of the
subtraction point $s_0$.
Intuitively, what the Omn\`es solution does is to correct a local weak
$K\to \pi\pi$ transition with an infinite chain of pion--loop bubbles,
incorporating the strong $\pi\pi\to\pi\pi$ rescattering to all orders
in $\chi$PT.
The Omn\`es exponential only sums a particular type of higher--order
Feynman diagrams, related to FSI. Therefore,
eq.~\eqn{eq:OMNES_WA} does not provide the complete result. 
Nevertheless, it allows us to perform a reliable estimate 
of higher--order effects because it does sum the most important
corrections.\footnote{
Indeed, a recent dispersive analysis of higher--order corrections induced
by $K\pi$ rescattering corroborates that these additional
``crossed--channel'' contributions are negligible \cite{BGKO:91}.}

The Omn\`es resummation of chiral logarithms is uniquely 
determined up to a polynomial (in $s$) ambiguity,
which has been solved with the large--$N_C$ amplitude
$\cA_I^{(R)\infty}$. The exponential only sums the
elastic rescattering of the final two pions, which is responsible
for the phase shift. Since the kaon mass is smaller than the
inelastic threshold, the virtual loop corrections from other 
intermediate states
($K\to K\pi, K\eta, \eta\eta, K\bar K \to\pi\pi$)
can be safely estimated at the one loop
level; they are included in ${\cal C}_I^{(R)}(s_0)$.

Taking the chiral prediction for $\delta^I_0(z)$ and expanding
$\Omega_I(M_K^2,s_0)$ to $\cO(p^2)$, 
\beq\label{eq:delta_omega}
\Omega_I(M_K^2,s_0) \;\approx\; 
1 +  {(M_K^2-s_0)\over\pi}\int
{dz\over (z-s_0)} {\delta^I_0(z)\over (z-M_K^2-i\epsilon)}
\;\equiv\; 1 + \delta\Omega_I(M_K^2,s_0) \, ,
\eeq
one should reproduce the one-loop $\chi$PT result.
This determines the factor ${\cal C}_I^{(R)}(s_0)$ to
$\cO(p^4)$ in the chiral expansion:
\beq
{\cal C}_I^{(R)}(s_0) \; = \; {\cal C}_I^{(R)}\left[ 1 -
\delta\Omega_I(M_K^2,s_0)\right] \;\approx\;
1 + \Delta_L\cA_I^{(R)} - \delta\Omega_I(M_K^2,s_0) \, .
\label{eq:ciszero}
\eeq
It remains a local ambiguity at higher orders
\cite{PP:00b,GP:97,BGKO:91}.

Eq.~\eqn{eq:OMNES_WA} allows us to improve the one-loop calculation,
by taking $s_0$ low enough that the $\chi$PT corrections
to ${\cal C}_I^{(R)}(s_0)$ are moderate and exponentiating the 
large logarithms with the Omn\`es factor.
Moreover, using the experimental phase shifts in the dispersive
integral one achieves an all--order resummation of FSI effects.
The numerical accuracy of this exponentiation has been successfully
tested \cite{PP:00b} through an analysis of the scalar pion form factor,
which has identical FSI than $A_0$.

At $s_0 =0$, the dispersive parts of the experimentally determined Omn\`es 
exponentials are
\cite{PP:00b}:
\beq\label{DISP}
\Re_0(M_K^2,0)  = 1.55 \pm 0.10\, ,
\qquad\qquad
\Re_2(M_K^2,0)  = 0.92 \pm 0.03\, .
\eeq
The quoted errors take into account uncertainties in the
experimental phase-shifts data and
additional inelastic contributions above the first inelastic threshold.
These numbers fit very well with the findings of the chiral one-loop
calculation discussed in the previous section.
The corrections induced by FSI in the moduli of the decay amplitudes
${\cal A}_I$ generate an enhancement of the
$\Delta I=1/2$ to $\Delta I=3/2$ ratio \cite{PP:00},
\beq
\Re_0(M_K^2,0)/\Re_2(M_K^2,0) = 1.68\pm 0.12 \, .\label{eq:ratio}
\eeq
This factor multiplies the enhancement already found at short distances.

At $\cO(p^4)$, the previous numbers should be corrected with 
the factors ${\cal C}_I^{(R)}(s_0)$,
which incorporate additional one-loop contributions not related to FSI.
These factors compensate the obvious $s_0$ dependence of the
Omn\`es exponentials, up to $\cO(p^6)$ corrections.
To estimate the remaining sensitivity to this parameter,
we have changed the subtraction point between $s_0=0$ and  
$s_0=3 M_\pi^2$ and have included the resulting fluctuations
in the final uncertainties. 
The detailed numerical analysis is given 
in appendix~\ref{sec:offshell}.
At $\nu=M_\rho$, we get the following
values for the resummed loop corrections:
\bea\label{eq:CI_MK}
\left|{\cal C}_0^{(8)}\right|  &= &
\Re_0(M_K^2,s_0)\; {\cal C}_0^{(8)}(s_0) \;\;\: = \; 1.31 \pm 0.06\, ,
\nonumber\\
\left|{\cal C}_0^{(27)}\right|  &= &
\Re_0(M_K^2,s_0)\; {\cal C}_0^{(27)}(s_0) \,\; =\; 2.4 \pm 0.1\, ,
\nonumber\\
\left|{\cal C}_0^{(ew)}\right|  &= &
\Re_0(M_K^2,s_0)\; {\cal C}_0^{(ew)}(s_0) \; =\; 1.31 \pm 0.07\, ,
\\
\left|{\cal C}_2^{(27)}\right|  &= &
\Re_2(M_K^2,s_0) \; {\cal C}_2^{(27)}(s_0) \,\; =\; 1.05 \pm 0.05\, .
\nonumber\\
\left|{\cal C}_2^{(ew)}\right|  &= &
\Re_2(M_K^2,s_0) \; {\cal C}_2^{(ew)}(s_0) \; =\; 0.62 \pm 0.05\, .
\nonumber
\eea

To derive the Omn\`es  representation, one makes use of Time--Reversal 
invariance, so that it can be strictly applied only to CP--conserving 
amplitudes. Nevertheless, the procedure can be directly extended to 
the CP--violating components relevant for the estimate of  
$\varepsilon^\prime /\varepsilon$.
Working to first order in the Fermi coupling, the CP--odd phase
is fully contained in the ratio of CKM matrix elements
$\tau$ which appears in the short--distance Wilson coefficients
and, therefore, in ${\cal A}_I^{(R)\infty}$.
Decomposing the isospin amplitudes
as ${\cal A}_I^{(R)} = {\cal A}_I^{(R)\, CP} 
+ \tau \,{\cal A}_I^{(R)\,\slashchar{CP}} $,
the Omn\`es solution can be derived separately
for the two amplitudes ${\cal A}_I^{(R)\, CP}$
 and ${\cal A}_I^{(R)\,\slashchar{CP}}$ 
 which respect Time--Reversal invariance.

\section{Numerical Analysis}
\label{sec:num}

The CP--violating ratio $\eps'/\eps$ is proportional to the CKM factor
Im($V^*_{ts} V^{\phantom{*}}_{td}$). The standard unitarity triangle 
analyses \cite{anamar} have estimated this parameter to be in the range
\beq\label{eq:Lt_num}
\mbox{\rm Im}(V^*_{ts} V^{\phantom{*}}_{td}) = 
(1.2\pm 0.2) \cdot 10^{-4} \, .
\eeq
This determination is obtained combining the present information on
various flavour--changing processes; mainly,
$\eps$, $B_0$--$\bar B_0$ mixing and the ratio 
$\Gamma(b\to u)/\Gamma(b\to c)$.
The final number is sensitive to the input values adopted for
several non-perturbative hadronic parameters and, thus,
there are large theoretical uncertainties \cite{PP:95} which are not easy
to quantify.

Since the Standard Electroweak Model has a unique source of
CP violation,
the same combination of CKM factors appears in the theoretical
prediction for $\eps$, which is proportional to the $K^0$--$\bar K^0$
matrix element of the $\Delta S=2$ operator:
\beq
\langle \bar{K}^0 |(\bar{s}_L \g_\mu d_L) (\bar{s}_L \g^\mu d_L)| K^0 \rangle
\equiv \fr{4}{3} f_K^2 M_K^2\, B_K(\mu) \ .
\eeq
The factor $B_K(\mu)$ parameterizes this hadronic matrix element in
vacuum insertion units. The corresponding Wilson coefficient 
$C_{\Delta S=2}(\mu)$ is known
at the next--to--leading logarithmic order \cite{BU:99,brev}. Taking
appropriate values for the different inputs one finds:
\beq
|\eps| \, = \, {4\over 3} \,\hat{B}_K \;
{\rm Im}\left(V^*_{ts} V^{\phantom{*}}_{td}\right) \; 
(18.9 - 14.4 \,\bar{\rho}) \, ,
\label{eq:epva}
\eeq
with $\bar{\rho}$ one of the two CKM parameters, in the Wolfenstein \cite{WO:83}
parameterization, which characterize the upper vertex of the unitarity
triangle. The standard analyses~\cite{anamar} favour the range 
$\bar{\rho}=0.2\pm 0.1$, implying
\beq\label{eq:eps_th}
|\eps |= \fr{4}{3}\, \hat{B}_K\; 
{\rm Im}\left(V^*_{ts} V^{\phantom{*}}_{td}\right) 
\; \left(16.0 \pm 1.4 \right) \ ,
\eeq
where $\hat{B}_K = C_{\Delta S=2}(\mu)\, B_K(\mu)$ 
is the scale--invariant bag parameter. In the large--$N_C$
limit, $\hat{B}_K = B_K(\mu) = 3/4$.

The numerical values of both ${\rm Im}\left(V^*_{ts} V^{\phantom{*}}_{td}\right)$
and $\bar{\rho}$ depend on hadronic inputs. However, $\eps$ is rather insensitive
to the precise value of $\bar{\rho}$; it changes by less than 10\%  when
$\bar{\rho}$ is varied within the previously quoted range.

Thus, we can make two different numerical analyses of $\eps'/\eps$:
\begin{enumerate}

\item The usual one, taking the experimental value of $\eps$ and adopting
the range \eqn{eq:Lt_num} for the relevant CKM factor.

\item Using instead the theoretical prediction of $\eps$ in eq.~\eqn{eq:eps_th},
the ratio $\eps'/\eps$ does not depend on
${\rm Im}\left(V^*_{ts} V^{\phantom{*}}_{td}\right)$ \cite{FA:00}. The
sensitivity of this CKM factor to different hadronic inputs is then reduced
to the explicit remaining dependence on $\hat{B}_K$.
\end{enumerate}

The second type of analysis is more suitable to a systematic $1/N_C$ approach.
The theoretical prediction for $\eps'/\eps$ depends on ratios of hadronic
matrix elements, i.e. $B_i/\hat{B}_K$.
It is known \cite{PP:95} that $\hat{B}_K$ has sizeable large--$N_C$ 
\cite{PR:91,RA:95,PER:00} and chiral~\cite{BSW:84} corrections,
which are of opposite sign and could then cancel to some extent.
Thus, one can expect the limit $N_C\to\infty$ to provide a good 
starting point to analyze
the relevant ratios $B_6^{(1/2)}/\hat{B}_K$ and $B_8^{(3/2)}/\hat{B}_K$.

We have performed the two types of numerical analysis, obtaining consistent
results. This allows us to estimate better the theoretical uncertainties, since
the two analyses have different sensitivity to hadronic inputs.
The contributions to Im($A_0$) from the operators $Q_{1,2,4}$ have been
estimated, following the strategy adopted in ref.~\cite{buras1}; i.e.
we have corrected them with the factor $\xi_0\approx 3.5$.

As a first estimate, we can  perform  the calculation of $\eps '/\eps$
to $\cO(p^4)$ in $\chi$PT, without making any Omn\`es resummation
of higher--order corrections. Once the large one-loop corrections
are taken into account, all important ingredients are already caught.
We find, for the two different types of analysis:
\beq
{\rm Re}(\eps'/\eps)\; =\; 
1.5 \cdot 10^{-3}\;\:
\fr{{\rm Im}\left(V^*_{ts} V^{\phantom{*}}_{td}\right)}{1.2 \cdot 10^{-4}} 
\; =\; 
1.8 \cdot 10^{-3}\, .
\label{eq:frk}
\eeq

To quantify the uncertainties, we need to consider higher--order effects.
Performing the Omn\`es resummation, as indicated in eq.~(\ref{eq:OMNES_WA}),
one finds:
\beq
\real\left(\eps'/\eps\right) \; =\;  
1.4 \cdot 10^{-3}\;\:
\fr{{\rm Im}\left(V^*_{ts} V^{\phantom{*}}_{td}\right)}{1.2 \cdot 10^{-4}} 
\; =\; 
 1.6 \cdot 10^{-3}\, .
\label{eq:frz}
\eeq
These numbers are quite close to the one-loop results \eqn{eq:frk}, 
which indicates
that the error induced by the chiral loop calculation is not large.

\indent From the previous numbers, we derive:
\beq
\real\left(\eps'/\eps\right) \; =\;  
\left(1.7\pm 0.2\, {}_{-0.5}^{+0.8} \pm 0.5\right) \cdot 10^{-3}\, .
\label{eq:final_result}
\eeq
The first error indicates the sensitivity to the short--distance
renormalization scale, which we have taken in the range
$M_\rho < \mu < m_c$.
The uncertainty coming from varying  the strange quark mass 
in the interval $(m_s+ m_q)(1\, \rm{GeV})=156\pm 25\, \rm{MeV}$
\cite{ms2000,PP:99,ALEPH:99,KM:00,JA:98,LatMass,BPR:98} is indicated 
by the second error.
We have added a 30\% uncertainty from
unknown next--to--leading in $1/N_C$ local contributions
(third error).

\section{Discussion}
\label{sec:summary}

The infrared effect of chiral loops generates an important enhancement
of the isoscalar $K\to\pi\pi$ amplitude. This effect gets amplified
in the prediction of $\varepsilon'/\varepsilon$, because 
at lowest order (in both $1/N_C$ and the chiral expansion) there
is an accidental numerical cancellation between the $I=0$ and $I=2$
contributions. Since the chiral loop corrections destroy this cancellation,
the final result for $\varepsilon'/\varepsilon$ is dominated by the
isoscalar amplitude. Thus, the Standard Model prediction for
$\varepsilon'/\varepsilon$ is finally governed by the matrix element
of the gluonic penguin operator $Q_6$.

There are three major ingredients in our theoretical analysis:
\begin{enumerate}

\item A short--distance calculation at the electroweak scale and
its renormalization--group evolution to the three--flavour theory
($\mu\lsim m_c$), which sums the leading ultraviolet logarithms.

\item The matching to the $\chi$PT description.

\item Chiral loop corrections, which induce large infrared logarithms
related to FSI.

\end{enumerate}

The first step is already known at the next--to--leading logarithmic 
order \cite{buras1,ciuc1}. The short--distance results are then very 
reliable.

We have tried to achieve an acceptable control of the large infrared 
chiral corrections, which are fully known at the one-loop level. 
A complete two-loop $\chi$PT calculation is not
yet available. Nevertheless, since the leading one-loop corrections 
are generated by the FSI of the two pions, we can use the Omn\`es
resummation to get an idea about the size to be expected for the
unknown higher--order contributions. The Omn\`es exponential 
only sums a particular type of higher--order Feynman diagrams,
related to FSI. Although it does not give the complete result,
it allows us to estimate the theoretical uncertainty
in a very reliable way, because it does sum the most important
corrections. The one-loop results and the Omn\`es calculation agree
within errors, indicating a good convergence of the chiral expansion.

The most critical step is the matching between the short-- and 
long--distance descriptions. 
We have performed this matching at leading
order in the $1/N_C$ expansion, where the result is exactly known
to $\cO(p^4)$ and $\cO(e^2p^2)$ in $\chi$PT
[$\cO(p^2)$ for $Q_6$].
This can be expected to provide a good approximation to the matrix 
elements of the leading $Q_6$ and $Q_8$ operators. Since all ultraviolet
and infrared logarithms have been resummed, our educated guess for
the theoretical uncertainty associated with $1/N_C$ corrections
is $\sim 30\%$.

As our final result we quote:
\beq
{\rm Re}(\eps'/\eps)\; =\; 
\left(1.7 \pm 0.9\right)\cdot 10^{-3}\; .
\eeq

A better determination of the strange quark mass would allow
to reduce the uncertainty to the 30\% level.
In order to get a more accurate prediction, it would be necessary to have
a good analysis of next--to--leading $1/N_C$ corrections. This is
a very difficult task, but progress in this direction can be
expected in the next few years \cite{trieste,BP:00,PR:91,PER:00,NLO_NC,latka}.

\section*{Acknowledgements}

We have benefited from discussions with G.~Colangelo, G.~Ecker,
M.~Knecht, J.~Portol\'es, J.~Prades and E.~de Rafael. 
This work has been supported by the European Union TMR Network 
``EURODAPHNE'' (Contract No. ERBFMX-CT98-0169) and by DGESIC, Spain
(Grant No. PB97-1261).

\appendix 
\renewcommand{\theequation}{\Alph{section}.\arabic{equation}} 
\setcounter{section}{0}\setcounter{equation}{0} 
  
\section{Octet basis transformation rules} 
\label{APP_D}
\setcounter{equation}{0} 

Following the same notation as the original references, 
one can change from the octet basis 
$\sum_i E_i\, O_i^8$ of ref.~\cite{BPP}
to the one of ref.~\cite{EKW:93}, $\sum_i N_i\, W_i^8$,
using either the following identities for the operators,
\beq
\begin{array}{rclrcl}
W_5^8\!\!\!&=&\!\!\!O_{10}^8 \, ; &  
W_{10}^8\!\!\!&=&\!\!\!O_1^8 \, ;   \\
W_6^8\!\!\!&=&\!\!\!\ds \fr{1}{2}O_{12}^8 \, ; & 
W_{11}^8\!\!\!&=&\!\!\!O_2^8 \, ;  \\
W_7^8\!\!\!&=&\!\!\!O_{13}^8 \, ; &  
W_{12}^8\!\!\!&=&\!\!\! O_3^8 \, ; \\
W_8^8\!\!\!&=&\!\!\!O_{10}^8+O_{11}^8
-\ds \fr{1}{2}\left(O_{12}^8+O_{13}^8\right) \, ;\qquad\qquad &
W_{13}^8\!\!\!&=&\!\!\!O_4^8 \, ;   \\
W_9^8\!\!\!&=&\!\!\!O_{15}^8 \, ;&  
W_{36}^8\!\!\!&=&\!\!\!O_1^8-O_3^8+O_5^8 \, ; 
\end{array}
\eeq
or the coefficient relations
\beq
\begin{array}{rclrcl}
N_5\!\!\!&=&\!\!\!E_{10}-E_{11} \, ; & 
N_{10}\!\!\!&=&\!\!\!E_1-E_5 \, ; \\
N_6\!\!\!&=&\!\!\!E_{11}+2 E_{12} \, ; &
N_{11}\!\!\!&=&\!\!\! E_2 \, ; \\
N_7\!\!\!&=&\!\!\!\ds \fr{1}{2} E_{11}+E_{13} \, ;
\qquad\qquad\qquad\qquad &
N_{12}\!\!\!&=&\!\!\! E_3+E_5 \, ; \\
N_8\!\!\!&=&\!\!\!E_{11} \, ;&  
N_{13}\!\!\!&=&\!\!\!E_4 \, ; \\
N_9\!\!\!&=&\!\!\!E_{15} \, ; &  
N_{36}\!\!\!&=&\!\!\!E_5\, .
\end{array}
\eeq

\section{ One-Loop Functions} 
\label{APP_B}
\setcounter{equation}{0}
 
The one-loop function $B(M_1^2,M_2^2,p^2)$ is defined by the
(dimensionally regularized) basic scalar integral with two bosonic
propagators:
\beq
i\int {d^D q\over (2\pi)^D}\, {1 \over
\left( M_1^2-q^2\right)\, \left[ M_2^2-(p-q)^2\right] }
\: = \: f_\pi^2\, B(M_1^2,M_2^2,p^2) +
\left\{ {2 \,\mu^{D-4}\over D-4} + \gamma_E - \ln{(4\pi)} - 1\right\} .
\eeq
It can be expressed in terms of the
function $\bar{J}_{12}(p^2)$ \cite{GL:85}: 
\beq 
f_\pi^2\; B(M_1^2,M_2^2,p^2)= -\bar{J}_{12}(p^2) + {1\over 16\pi^2}\left ( 
 \ln  \frac{M_2^2}{\nu^2} + {M_1^2\over M_1^2-M_2^2} 
\ln {M_1^2\over M_2^2} \right )\, ,
\eeq 
with 
\bea
\bar{J}_{12}(p^2) &=& \frac{1}{16\pi^2}\left\{ 1 -\frac{1}{2}
\left ( 1+\frac{M_1^2}{p^2}
-\frac{M_2^2}{p^2}\right )\;\ln\frac{M_1^2}{M_2^2}
 +\frac{M_1^2}{M_1^2-M_2^2}\;\ln\frac{M_1^2}{M_2^2} 
 \right .\nonumber\\ &&\qquad\left.\mbox{} 
-\frac{1}{2}\frac{\lambda}{p^2}\;
 \ln \frac{(p^2+\lambda)^2-(M_1^2-M_2^2)^2}
{(p^2-\lambda)^2-(M_1^2-M_2^2)^2} \right \}\, ,
\eea
where
\beq
\lambda^2 \equiv \lambda^2(p^2,M_1^2,M_2^2)
= \left [p^2-(M_1+M_2)^2\right ] \left [p^2-(M_1-M_2)^2\right ]\, .
\eeq
For $M_1=M_2\equiv M$ one gets 
\beq \label{eq:Beqmasses}
f_\pi^2\; B(M^2,M^2,p^2)= -\bar{J}(p^2) + {1\over 16\pi^2}\left ( 
 \ln  \frac{M^2}{\nu^2} + 1 \right )\, , 
\eeq 
where $\bar{J}(p^2)$ is given by
\beq 
\bar{J}(p^2) = {1\over (4\pi )^2}\left\{ 2 - \sigma\ln{\left ( 
{\sigma +1\over \sigma -1}\right )}\right\} 
\qquad ; \qquad 
\sigma\equiv\sqrt{1-{4 M^2\over p^2}}\, . 
\label{eq:Jb_PP} 
\eeq 

The one-loop amplitudes contain an additional logarithmic dependence on 
the chiral renormalization scale $\nu$, through the factors
 \ ($ P=\pi ,\, K,\, \eta$):
\beq 
\nu_P = {M_P^2\over 32\pi^2 f_\pi^2} \;\ln {M_P^2\over \nu^2}\, . 
\eeq 
%

\section{One-loop amplitudes}
\label{sec:onshell}\setcounter{equation}{0}

The one-loop $K\to\pi\pi$ amplitudes have been computed in
refs.~\cite{KA91,BPP},
in the absence of electroweak corrections.
The electromagnetic contributions have been recently calculated
in refs.~\cite{EIMNP:00,CDG:99,CG:00}. The results take the form:
\bea 
\label{ONELOOP_8a} 
\Delta_L {\cal A}_0^{(8)} &=&
 \Biggl\{ 
-{1\over 2}\left (2 M_K^2-M_\pi^2\right ) B(M_\pi^2,M_\pi^2,M_K^2)
-{1\over 18}M_\pi^2 B(M_\eta^2,M_\eta^2,M_K^2)
\nonumber\\
&&\;\;\mbox{}
+{1\over 4}{M_K^2\over M_\pi^2}\left ( M_K^2-4M_\pi^2\right )
B(M_K^2,M_\pi^2,M_\pi^2)
+{1\over 12}{M_K^4\over M_\pi^2}B(M_K^2,M_\eta^2,M_\pi^2)
\nonumber\\
&& \;\;\mbox{}
+\fr{M_K^4}{4 (M_K^2-M_\pi^2) M_\pi^2}\left[
 \left(2+15\fr{M_\pi^2}{M_K^2} -21 \fr{M_\pi^4}{M_K^4}\right)\nu_\pi
\right. \nn \\
& & \left.\;\;\mbox{}
+2 \fr{M_\pi^2}{M_K^2} \left(3  + \fr{M_\pi^2}{M_K^2} \right)\nu_K 
+\left(-2+3\fr{M_\pi^2}{M_K^2}-5\fr{M_\pi^4}{M_K^4}\right)\nu_\eta
\right]
\Biggr\}\, ,
\\
\label{ONELOOP_8b}
\Delta_L{\cal A}_{0}^{(ew)} &=& \Biggl\{ 
-{1\over 2}\left (2 M_K^2-M_\pi^2\right ) B(M_\pi^2,M_\pi^2,M_K^2)
+\fr{3 M_K^2}{8} B(M_K^2,M_K^2,M_K^2)
\nn \\ & & \;\;\mbox{}
+{1\over 4}{M_K^2\over M_\pi^2}\left ( M_K^2-4M_\pi^2\right )
B(M_K^2,M_\pi^2,M_\pi^2)
+\fr{M_K^4}{8 M_\pi^2} B(M_K^2,M_\eta^2,M_\pi^2)
\nn \\ & & \;\;\mbox{}
+\fr{1}{4}\left(17+2\fr{M_K^2}{M_\pi^2}\right)
\nu_\pi
+\fr{1}{4} \left(4+\fr{M_K^2}{M_\pi^2}\right) \nu_K 
+\fr{3}{4}\left(1-\fr{M_K^2}{M_\pi^2}\right)
\nu_\eta
 \Biggr\}\, , \qquad
\eea 
for the octet isoscalar amplitude,   
\bea 
\label{ONELOOP_27a} 
\Delta_L {\cal A}_0^{(27)} &=&\Biggl\{ 
-{1\over 2}\left (2 M_K^2-M_\pi^2\right ) B(M_\pi^2,M_\pi^2,M_K^2) 
+{1\over 2}M_\pi^2 B(M_\eta^2,M_\eta^2,M_K^2)
\nonumber\\
&&\;\;\mbox{}
+{1\over 4}{M_K^2\over M_\pi^2}\left ( M_K^2-4M_\pi^2\right )
B(M_K^2,M_\pi^2,M_\pi^2)
-{1\over 3}{M_K^4\over M_\pi^2}B(M_K^2,M_\eta^2,M_\pi^2)
\nonumber\\
&& \;\;\mbox{}
+ \fr{M_K^4}{4 (M_K^2-M_\pi^2) M_\pi^2} 
\left[ 
\left(2+15\fr{M_\pi^2}{M_K^2}-21 \fr{M_\pi^4}{M_K^4}\right)\nu_\pi
\right.
\nn \\
& & \left.\;\;\mbox{}-
\left(10+4\fr{M_\pi^2}{M_K^2}-22 \fr{M_\pi^4}{M_K^4}\right)\nu_K+
\left(8-27\fr{M_\pi^2}{M_K^2}+15 \fr{M_\pi^4}{M_K^4}\right)\nu_\eta
\right]
 \Biggr\}\, , \quad\quad\;
\eea 
for the 27--plet isoscalar amplitude and 
\bea 
\label{ONELOOP_2a} 
\Delta_L{\cal A}_2^{(27)} &=&
\Biggl\{ 
{1\over 2}\left ( M_K^2-2M_\pi^2\right ) B(M_\pi^2,M_\pi^2,M_K^2) 
\nonumber\\ &&\;\;\mbox{}
+{5\over 8}{M_K^2\over M_\pi^2}\left (M_K^2-{8\over 5}M_\pi^2\right )
B(M_K^2,M_\pi^2,M_\pi^2)
+{1\over 24}{M_K^4\over M_\pi^2}B(M_K^2,M_\eta^2,M_\pi^2) 
\nonumber\\
&&\;\;\mbox{}
 +\fr{M_K^4}{4 (M_K^2-M_\pi^2) M_\pi^2} \left[
\left(5 -18 \fr{M_\pi^2}{M_K^2}+21 \fr{M_\pi^4}{M_K^4} \right)\nu_\pi 
\right. \nn \\
& & \left.\;\;\mbox{}
-\left( 4 -2 \fr{M_\pi^2}{M_K^2}+ 2\fr{M_\pi^4}{M_K^4} \right)\nu_K
-
\left( 1+ 3 \fr{M_\pi^4}{M_K^4} \right)\nu_\eta
\right]\Biggr\} \, ,
\\
\label{ONELOOP_2b}
\Delta_L{\cal A}_{2}^{ew}
&=& \Biggl\{
{1\over 2}\left ( M_K^2-2M_\pi^2\right ) B(M_\pi^2,M_\pi^2,M_K^2) 
\nonumber\\ &&\;\;\mbox{}
+{5\over 8}{M_K^2\over M_\pi^2}\left (M_K^2-{8\over 5}M_\pi^2\right )
B(M_K^2,M_\pi^2,M_\pi^2)
+ \fr{M_K^4}{8 M_\pi^2} B(M_K^2,M_\eta^2,M_\pi^2)
\nn \\ & & \;\;\mbox{}
+\fr{5}{4}\left( \fr{M_K^2}{M_\pi^2}+\fr{11}{5}\right)\nu_\pi 
-\fr{1}{2}\left(\fr{M_K^2}{M_\pi^2}-5\right)\nu_K
-\fr{3}{4}\left(\fr{M_K^2}{M_\pi^2}-1\right)\nu_\eta
\Biggr\}\, , \qquad
\eea
for the $I=2$ amplitude.

\section{Resummation of higher--order corrections}
\label{sec:offshell}
\setcounter{equation}{0} 
In this appendix we provide some details on the Omn\`es procedure for
calculating the isospin amplitudes.
The resummed loop corrections are contained in 
the factors ${\cal C}_I^{(R)}$, as defined in eq.~\eqn{eq:OMNES_WA}.
%
%
At $\cO(p^4)$ in the chiral expansion, these quantities should reproduce
the one-loop $\chi$PT results in \eqn{eq:onel}; this determines
the factors ${\cal C}_I^{(R)}(s_0)$, with $s_0$ the subtraction point,
up to higher--order local contributions:
\beq
{\cal C}_I^{(R)}(s_0) \; = \; {\cal C}_I^{(R)}\left[ 1 -
\delta\Omega_I(M_K^2,s_0)\right] \;\approx\;
1 + \Delta_L\cA_I^{(R)} - \delta\Omega_I(M_K^2,s_0) \, .
\eeq
Here, $\Delta_L\cA_I^{(R)}$ is the one-loop $\chi$PT result and 
$\delta\Omega_I(M_K^2,s_0)$ is obtained by 
taking the chiral prediction for the phase shift $\delta^I_0(z)$ in 
$\Omega_I(M_K^2,s_0)$ and expanding
$\Omega_I(M_K^2,s_0)$ to $\cO(p^2)$, 
\beq
\Omega_I(M_K^2,s_0) \; = \; 1 + \delta\Omega_I(M_K^2,s_0) + \cO(p^4)\, .
\eeq
The explicit expressions for $\Delta_L\cA_I^{(R)}$ are listed in appendix 
\ref{sec:onshell}. The once-subtracted  Omn\`es exponential 
\beq
\Omega_I(M_K^2,s_0)\; =\;
 \exp{\left\{ {(M_K^2-s_0)\over\pi}\int_{4 M_\pi^2}^{\bar{z}}
{dz\over (z-s_0)} {\delta^I_0(z)\over (z-M_K^2-i\epsilon)}\right\}}
\eeq
contains the integral over the experimentally determined phase shift 
$\delta^I_0(z)$ with $I=0$ or $2$. The upper edge of the integral $\bar{z}$ 
should correspond to the first inelastic threshold in the given isospin 
channel.
The corresponding expansion factor, 
\beq
\delta\Omega_I(M_K^2,s_0)\; =\;
  {(M_K^2-s_0)\over\pi}\int_{4 M_\pi^2}^{\bar{z}}
{dz\over (z-s_0)} {\delta^I_0(z)\over (z-M_K^2-i\epsilon)} \, ,
\eeq
contains the same dispersive integral, but with the phase shift  
$\delta^I_0(z)$ determined at $\cO(p^2)$ in $\chi$PT.
The explicit expressions for $\delta\Omega_I(M_K^2,s_0)$ with $I=0$ and 2 are 
as follows:
\bea
\delta \Omega_0(M_K^2,s_0)\!\!&=&\!\! \fr{1}{32\,\pi^2 f_\pi^2}\left\{
\left(2 \, M_K^2-M_\pi^2\right) \,\s(M_K^2) \,
\ln{\left[\fr{\s(M_K^2)-\s(\bar z )}{\s(M_K^2)+\s(\bar z )}\right]}
\right.\nn \\ 
& &\left. -\left(2 \, s_0-M_\pi^2\right)\, \s(s_0)\, 
\ln{\left[\fr{\s(s_0)-\s(\bar z )}{\s(s_0)+\s(\bar z )}\right]} 
 - 2\,\left(M_K^2-s_0\right)\,
\ln{\left[\fr{1-\s(\bar z )}{1+\s (\bar z )}\right]}
\right\} \, ,
\nn\\ &&\\
\delta \Omega_2(M_K^2,s_0)\!\!&=&\!\!\fr{1}{32\,\pi^2 f_\pi^2}\left\{
\left(2 \, M_\pi^2-M_K^2\right)\,\s(M_K^2) \,
\ln{\left[\fr{\s(M_K^2)-\s(\bar z )}{\s(M_K^2)+\s(\bar z )}\right]}
\right.\nn \\ & & \left.
-\left(2 \, M_\pi^2-s_0\right)\,\s(s_0) \,
\ln{\left[\fr{\s(s_0)-\s(\bar z )}{\s(s_0)+\s(\bar z )}\right]} 
+\left(M_K^2-s_0\right)\,
\ln{\left[\fr{1-\s (\bar z )}{1+\s ( \bar z )}\right]}\right\}\, ,
\nn\eea
where for convenience we have defined $\s(s)\equiv \sqrt{1-4 M_\pi^2/s}$.

In the following numerical analysis we have varied the subtraction 
point between $s_0=0$ and $s_0=3M_\pi^2$, 
together with the upper edge of the Omn\`es integral $\bar{z}$,
to estimate the sensitivity of our predictions to these parameters.
We have fixed the $\chi$PT renormalization scale at $\nu=M_\rho$. 
In tables \ref{tab:mat11} and \ref{tab:mat12} the dispersive part of the 
Omn\`es factors and $\delta \Omega_I(M_K^2,s_0)$ are reported as functions of
$s_0$, for $\bar{z}=1$ GeV${}^2$ and 
$\bar{z}=2$ GeV${}^2$ respectively.
%
\begin{table}[!ht] 
\begin{center}
\caption{The $s_0$ dependence of the once-subtracted Omn\`es parameters 
for $\bar{z}=1$ GeV$^2$. } 
\label{tab:mat11} 
\vspace{0.3cm}
\begin{tabular}{|c|c|c||c|c|}\hl \hl 
 $s_0$ &  $\Re_0 $ & $\delta \Omega_0$ &$\Re_2 $ & $\delta \Omega_2$  \\ \hl 
  $  0$ & 1.45 & 0.32 + 0.46 i & 0.94 & -0.16 - 0.20 i \\ 
  $ M_\pi^2$ & 1.40 & 0.29 + 0.46 i & 0.95 & -0.15 - 0.20 i \\ 
  $ 2 M_\pi^2$ & 1.33 & 0.25 + 0.46 i & 0.96 & -0.13 - 0.20 i \\ 
  $ 3 M_\pi^2$ & 1.26 & 0.21 + 0.46 i & 0.97 & -0.12 - 0.20 i \\ 
\hl \hl 
\end{tabular} 
\end{center}
\end{table}
\begin{table}[!ht] 
\begin{center} 
\caption{The $s_0$ dependence of the once-subtracted Omn\`es parameters 
for  $\bar{z}=2$ GeV$^2$. } 
\label{tab:mat12}
\vspace{0.3cm}
\begin{tabular}{|c|c|c||c|c|}\hl \hl 
 $s_0$ &  $\Re_0 $ & $\delta \Omega_0$ &$\Re_2 $ & $\delta \Omega_2$  \\ \hl 
  $  0$ & 1.58 & 0.47 + 0.46 i & 0.92 & -0.24 - 0.20 i \\ 
  $ M_\pi^2$ & 1.51 & 0.43 + 0.46 i & 0.93 & -0.22 - 0.20 i \\ 
  $ 2 M_\pi^2$ & 1.44 & 0.39 + 0.46 i & 0.94 & -0.20 - 0.20 i \\ 
  $ 3 M_\pi^2$ & 1.35 & 0.33 + 0.46 i & 0.95 & -0.17 - 0.20 i \\ 
\hl \hl 
\end{tabular}
\end{center}
\end{table}
\begin{table} [!ht] 
\begin{center}
\caption{Resummed loop corrections with one subtraction
and $\bar{z}=1$ GeV$^2$. } 
\label{tab:mat13}
\vspace{0.3cm}
\begin{tabular}{|c|c|c|c|c|c|}\hl \hl 
$s_0$ & $ |{\cal C}_0^{(8)}|$   & $ |{\cal C}_0^{(27)}|$  & 
$ |{\cal C}_0^{(ew)}|$ & $ |{\cal C}_2^{(27)}|$  & $ |{\cal C}_2^{(ew)}|$  
\\ \hl 
  $  0$ & 1.37 & 2.47 & 1.38 & 1.06 & 0.62 \\ 
 \hl $ M_\pi^2$ & 1.36 & 2.42 & 1.37 & 1.05 & 0.61 \\ 
 \hl $ 2 M_\pi^2$ & 1.35 & 2.36 & 1.36 & 1.05 & 0.60 \\ 
 \hl $ 3 M_\pi^2$ & 1.33 & 2.28 & 1.34 & 1.04 & 0.59 \\ 
 \hl\hl \hl 
\end{tabular}
\end{center} 
\end{table}
\begin{table}[!ht] 
\begin{center}
\caption{Resummed loop corrections with one subtraction
and $\bar{z}=2$ GeV$^2$. } 
\label{tab:mat14}
\vspace{0.3cm}
\begin{tabular}{|c|c|c|c|c|c|}\hl \hl 
$s_0$ & $ |{\cal C}_0^{(8)}|$  & $ |{\cal C}_0^{(27)}|$   & 
$ |{\cal C}_0^{(ew)}|$   & $ |{\cal C}_2^{(27)}|$ & $ |{\cal C}_2^{(ew)}|$ 
 \\ \hl 
 $  0$ & 1.26 & 2.45 & 1.27 & 1.10 & 0.68 \\ 
 $ M_\pi^2$ & 1.27 & 2.41 & 1.27 & 1.10 & 0.67 \\ 
 $ 2 M_\pi^2$ & 1.27 & 2.36 & 1.28 & 1.09 & 0.65 \\ 
 $ 3 M_\pi^2$ & 1.26 & 2.28 & 1.27 & 1.08 & 0.64 \\ 
\hl \hl 
\end{tabular}
\end{center} 
\end{table}
The corresponding moduli of the corrections
$\,{\cal C}_I^{(R)}$, derived according to 
eq.~\eqn{eq:CI_MK}, are given in tables \ref{tab:mat13} and \ref{tab:mat14}.
The residual tiny dependence of $\, |{\cal C}_I^{(R)}|$ 
on the subtraction point $s_0$ should be cancelled by 
missing $\cO(p^6)$ contributions to $\, {\cal C}_I^{(R)}(s_0)$,
since the local ambiguity of the Omn\`es procedure has been only solved 
to $\cO(p^4)$ in the chiral expansion. From 
tables \ref{tab:mat13} and \ref{tab:mat14} one can also verify that 
the once-subtracted result is sensitively dependent on  $\bar{z}$.

\begin{table}[!ht] 
\begin{center}
\caption{The $\bar z$ dependence of the twice-subtracted Omn\`es parameters 
for $s_0=0$.}
\label{tab:mat21}
\vspace{0.3cm}
\begin{tabular}{|c|c|c||c|c|}\hl \hl 
 $\bar z$ (GeV$^2$) &  $\Re_0 $ & $\delta \Omega_0$ &$\Re_2 $ & 
$\delta \Omega_2$  \\ \hl 
  $  1$ & 1.44 & 0.40 + 0.46 i & 0.86 & -0.20 - 0.20 i \\ 
  $  2$ & 1.46 & 0.42 + 0.46 i & 0.85 & -0.21 - 0.20 i \\ 
 \hl \hl 
\end{tabular}
\end{center} 
\end{table}
\begin{table}[!ht] 
\begin{center}
\caption{Resummed loop corrections with two subtractions and $s_0=0$.} 
\label{tab:mat23}
\vspace{0.3cm}
\begin{tabular}{|c|c|c|c|c|c|}\hl \hl 
$\bar z$ (GeV$^2$) & $ |{\cal C}_0^{(8)}|$  & $ |{\cal C}_0^{(27)}|$ & 
$ |{\cal C}_0^{(ew)}|$ & $ |{\cal C}_2^{(27)}|$  & $ |{\cal C}_2^{(ew)}|$ 
 \\ \hl 
  $  1$ & 1.25 & 2.34 & 1.26 & 1.00 & 0.60 \\ 
  $  2$ & 1.23 & 2.34 & 1.24 & 1.00 & 0.61 \\ 
\hl \hl 
\end{tabular}
\end{center}
\end{table}

As it was noticed in ref.~\cite{PP:00b},
the sensitivity to the higher energy region of the dispersive integral 
(i.e. the numerical  dependence on the upper edge $\bar{z}$) 
is reduced by performing more subtractions.
However, a  better knowledge of the theory is required in this case.
Indeed, the sensitivity to unknown higher--order corrections in the chiral 
expansion will increase with the number of subtractions, so that 
the resulting amplitudes can only be trusted
at the lowest values of the subtraction point ($s_0\sim 0$), 
where $\chi$PT corrections are moderate.
We have checked these statements using the twice-subtracted 
Omn\`es exponential \cite{PP:00b}:
\beq
\Omega_I(M_K^2,s_0) =
 \exp{\left\{ (M_K^2-s_0)\,\fr{g^\prime_I(s_0)}{1+g_I(s_0)} + 
{(M_K^2-s_0)^2\over\pi}\int_{4 M_\pi^2}^{\bar{z}}
{dz\over (z-s_0)^2} {\delta^I_0(z)\over (z-M_K^2-i\epsilon)}\right\}}\, ,
\eeq
where the functions $g_I(s)$ (and their first derivatives $g^\prime_I(s)$)
are the one-loop contributions (and their derivatives) to the isospin
amplitudes due to the elastic $\pi\pi$ rescattering:
\bea
g_0(s)&=&-\fr{1}{2} (2 \, s -M_\pi^2)\;B(M_\pi^2,M_\pi^2,s)\, ,
\nn \\ 
g_2(s)&=&\phantom{-}\fr{1}{2} ( s- 2\, M_\pi^2)\;B(M_\pi^2,M_\pi^2,s)\, .
\eea
The expansion of $\Omega_I(M_K^2,s_0)$ at $\cO(p^2)$ defines
\beq
\delta \Omega_I(M_K^2,s_0)\; =\; (M_K^2-s_0)\, g^\prime_I(s_0) +
\fr{(M_K^2-s_0)^2}{\pi}\int_{4 M_\pi^2}^{\bar z}
         \fr{{\rm d}z}{(z-s_0)^2}\fr{\delta_I(z)}{(z-M_K^2-i\eps)}\, ,
\eeq
where the phase shift $\delta_0^I(z)$ is taken at $\cO(p^2)$ in $\chi$PT.
The numerical results obtained at $s_0=0$, with the twice-subtracted
Omn\`es procedure, are reported 
in tables \ref{tab:mat21} and \ref{tab:mat23}.

The final results for the moduli of the 
correction factors ${\cal C}_I^{(R)}$,
quoted in eq.~\eqn{eq:CI_MK},
take into account the sensitivity to $s_0$ and $\bar{z}$
of the once-subtracted solution and the values obtained
at $s_0 =0$ with two subtractions.



\begin{thebibliography}{99}

\bibitem{PP:00} E. Pallante and A. Pich,  Phys. Rev. Lett.
    84 (2000) 2568.    

\bibitem{PP:00b} E. Pallante and A. Pich, Nucl. Phys. B592 (2000) 294.

\bibitem{na48}
  A. Ceccucci, {\it New measurement of direct CP
  violation in two pion decays of neutral kaons by experiment
  NA48 at CERN}, CERN Particle Physics Seminar (February 29, 2000),
  http://www.cern.ch/NA48/Welcome.html;
  NA48 collaboration (V. Fanti {\em et al.}), Phys. Lett. B465
  (1999) 335.  

\bibitem{ktev} KTeV collaboration (A. Alavi--Harati {\em et al.}) 
               \prl  83 (1999) 22.

\bibitem{munich}
 A.J. Buras et al., Nucl. Phys. B592 (2001) 55;  
 S.~Bosch et al., Nucl. Phys. B565 (2000) 3;   
 G.~Buchalla et al., Rev. Mod. Phys. 68 (1996) 1125;
 A.J.~Buras et al., Nucl. Phys. B408 (1993) 209,
 B400 (1993) 37, 75, B370 (1992) 69.

\bibitem{rome}
  M. Ciuchini et al., Nucl. Phys. B (Proc. Suppl.) 99 (2001) 27; 
  hep-ph/9910237;
  Nucl. Phys. B523 (1998) 501, B415 (1994) 403;
  Phys. Lett. B301 (1993) 263.

\bibitem{trieste}
 S.~Bertolini et al., Rev. Mod. Phys. 72 (2000) 65; 
 Phys. Rev. D63 (2001) 056009;   
 Nucl. Phys. B449 (1995) 197, B476 (1996) 225, B514 (1998) 63, 93.

\bibitem{FA:00}  M.~Fabbrichesi, Phys. Rev. D62 (2000) 097902.

\bibitem{dortmund}
 T. Hambye et al.,  hep-ph/0001088;    
 Nucl. Phys. B564 (2000) 391;
 Eur. Phys. J. C10 (1999) 271;
 Phys. Rev. D58 (1998) 014017;
 Y.-L. Wu, hep-ph/0012371.

\bibitem{BP:00} J. Bijnens and J. Prades, JHEP 06 (2000) 035;
  Nucl. Phys. (Proc. Suppl.) 96 (2001) 354. 

\bibitem{we:00} A.~Pich, Nucl. Phys. B (Proc. Suppl.) 93 (2001) 253;
  E. Pallante, A.~Pich and I. Scimemi, hep-ph/0010229; hep-ph/0010073.

\bibitem{dubna} A.A. Belkov et al., hep-ph/9907335.

\bibitem{taipei} H.Y. Chen, Chin. J. Phys. 38 (2000) 1044.  

\bibitem{NA:00} S. Narison, Nucl. Phys. B593 (2001) 3. 

\bibitem{beyond} There is a vast literature on this issue,
which can be traced back from the recent review by
A. Masiero and O. Vives, hep-ph/0104027. 

\bibitem{tHO:74} G. 't Hooft, \np B72 (1974) 461; B75 (1974) 461.

\bibitem{WI:79} E. Witten, \np B149 (1979) 285; B160 (1979) 57;
  Ann. Phys. 128 (1980) 363.

\bibitem{BBG87} W.A. Bardeen, A.J. Buras and J.-M. G\'erard,
  Nucl. Phys. B293 (1987) 787; Phys. Lett. B211 (1988) 343,
  B192 (1987) 138, B180 (1986) 133;
  A.J. Buras and J.-M.~G\'erard, Nucl. Phys. B264 (1986) 371.

\bibitem{WI:69} K.G. Wilson \pr 179 (1969) 1499; 
  W. Zimmermann, {\it Lectures on Elementary Particles and Quantum
  Field Theory}, Brandeis Summer Institute (1970) Vol.~1 (MIT Press,
  Cambridge, MA, 1970).
  
\bibitem{RGroup}
  E.C.G. Stueckelberg and A. Peterman, Helv. Phys. Acta 26 (1953) 499;
  M. Gell-Mann and F.E. Low, Phys. Rev. 95 (1954) 1300;
  C. Callan Jr., Phys. Rev. D2 (1970) 1541;
  K. Symanzik, Commun. Math. Phys. 18 (1970) 227; 23 (1971) 49.

\bibitem{EFT} A. Pich, {\it Effective Field Theory}, in
  ``Probing the Standard Model of Particle Interactions'', Proc.
  1997 Les Houches Summer School, eds. R. Gupta {\em et al.}
  (Elsevier, Amsterdam, 1999), Vol. II, p.~949.

\bibitem{GLAM:74} 
  M.K. Gaillard and B.W. Lee, Phys. Rev. Lett. 33 (1974) 108;
  G. Altarelli and L.~Maiani, Phys. Lett. B52 (1974) 351. 

\bibitem{VZS:75} A.I. Vainshtein, V.I. Zakharov and M.A. Shifman,
  JETP Lett. 22 (1975) 55; Nucl. Phys. B120 (1977) 316.

\bibitem{GW:79} F.J. Gilman and M.B. Wise, \pr D20 (1979) 2392;
       \pr D21 (1980) 3150.

\bibitem{BU:99} A.J.~Buras, {\it Weak Hamiltonian, CP Violation and
   Rare Decays}, in
  ``Probing the Standard Model of Particle Interactions'', Proc.
  1997 Les Houches Summer School, eds. R.~Gupta {\em et al.}
  (Elsevier, Amsterdam, 1999), Vol. I, p.~281.

\bibitem{buras1} A.J.~Buras, M.~Jamin and  M.E.~Lautenbacher,
   \np B408 (1993) 209; \pl  B389 (1996) 749.

\bibitem{ciuc1} M.~Ciuchini {\em et al.}, \pl { B301} (1993) 263;
    Z. Phys. C68 (1995) 239.

\bibitem{WE:79} S. Weinberg, Physica 96A (1979) 327.

\bibitem{GL:85} J. Gasser and H. Leutwyler, Ann. Phys., NY 158 (1984) 142;
     Nucl. Phys. B250 (1985) 456; 517; 539.

\bibitem{EC:95} G. Ecker, Prog. Part. Nucl. Phys. 35 (1995) 1.

\bibitem{PI:95} A. Pich, Rep. Prog. Phys. 58 (1995) 563.

\bibitem{ME:93} U.-G. Meissner, Rep. Prog. Phys. 56 (1993) 903.

\bibitem{GM:91} J. Gasser, U-G. Meissner, Phys. Lett. B258 (1991) 219. 

\bibitem{Omega} J.F. Donoghue et al., Phys. Lett. B179 (1986) 361;
  H.Y. Cheng, Phys. Lett. B201 (1988) 155;
  M. Lusignoli, Nucl. Phys. B325 (1989) 33.

\bibitem{BG:87}
  A.J.~Buras and J.-M. G\'erard, Phys. Lett. B192 (1987) 156.

\bibitem{EMNP:00} G. Ecker, G. M\"uller, H. Neufeld and A. Pich,
    Phys. Lett. B477 (2000) 88.

\bibitem{CR:67} J. A. Cronin, \pr 161 (1967) 1483.

\bibitem{BW:84} J. Bijnens and M.B. Wise, \pl B137 (1984) 245.

\bibitem{GRW:86} B. Grinstein, S.-J. Rey and M.B. Wise, Phys. Rev. D33
   (1986) 1495.

\bibitem{BE:85} C. Bernard et al., Phys. Rev. D32 (1985) 2343.

\bibitem{CR:86} R.J. Crewther, Nucl. Phys. B264 (1986) 277.

\bibitem{KA91}
  J. Kambor, J. Missimer and D. Wyler, Nucl. Phys. B346 (1990) 17;
  Phys. Lett. B261 (1991) 496; 
  J. Kambor et al., Phys. Rev. Lett. 68 (1992) 1818.

\bibitem{BPP}
  J. Bijnens, E. Pallante and J.~Prades, Nucl. Phys. B521 (1998) 305;
  E. Pallante, JHEP 01 (1999) 012.


\bibitem{PGR:86} A. Pich, B. Guberina and E. de Rafael, \np B277 (1986)
    197.

\bibitem{EIMNP:00} G. Ecker, G. Isidori, G. M\"uller, H. Neufeld
   and A. Pich, Nucl. Phys. B591 (2000) 419;  
   work in progress.  

\bibitem{CDG:99} V. Cirigliano, J.F. Donoghue and E. Golowich,
   Phys. Lett. B450 (1999) 241; Phys. Rev. D61 (2000) 093001, 093002.

\bibitem{CG:00} V. Cirigliano and E. Golowich, Phys. Lett. B475 (2000) 351.

\bibitem{EKW:93} G. Ecker, J. Kambor and D. Wyler, Nucl. Phys. B394
   (1993) 101.

\bibitem{dR:89} E. de Rafael, Nucl. Phys. B (Proc. Suppl.) 7A (1989) 1.

\bibitem{PI:89} A. Pich, Nucl. Phys. B (Proc. Suppl.) 7A (1989) 194.

\bibitem{PR:91} A. Pich and E. de Rafael, Nucl. Phys. B358 (1991) 311;
  Phys. Lett. 374 (1996) 186.

\bibitem{JP:94} M. Jamin and A. Pich, Nucl. Phys. B425 (1994) 15.

\bibitem{EGPR:89} 
 G. Ecker, J. Gasser, A. Pich and E. de Rafael, Nucl. Phys. B321 (1989) 
 311; G. Ecker, J. Gasser, H. Leutwyler, A. Pich and E. de Rafael, 
 Phys. Lett. B223 (1989) 425.

\bibitem{PPR:98} 
   S. Peris, M. Perrotet and E. de Rafael, JHEP 05 (1998) 011.

\bibitem{JOP:00}
   M. Jamin, J.A. Oller and A. Pich, Nucl. Phys. B587 (2000) 331;
   work in progress.

\bibitem{ABT:00}
   G. Amor\'os, J. Bijnens and P. Talavera, Phys. Lett. B480 (2000) 71;
     Nucl. Phys. B585 (2000) 293.  

\bibitem{BCE:99} J. Bijnens, G. Colangelo, G. Ecker, JHEP 02 (1999) 020.

\bibitem{ms2000} 
M. Davier, S. Chen, A. H\"ocker, J. Prades and A. Pich, 
{\it Strange quark mass from $\tau$ decays},
Proc. 6th Int. Workshop on Tau Lepton Physics (Victoria, 
Canada, 2000), Nucl. Phys. B (Proc. Suppl.) 98 (2001).

\bibitem{PP:99}
A. Pich and J.~Prades, JHEP 10 (1999) 004, JHEP 06 (1998) 013.

\bibitem{ALEPH:99}
ALEPH Collab., Eur. Phys. J. C11 (1999) 599, C10 (1999) 1.

\bibitem{KM:00}
J. Kambor and K. Maltman, Phys. Rev. D62 (2000) 093023;
K.G. Chetyrkin et al., Nucl. Phys. B533 (1998) 473.

\bibitem{JA:98} 
  M. Jamin, Nucl. Phys. B (Proc. Suppl.) 64 (1998) 250;
  K.G. Chetyrkin et al., Phys. Lett. B404 (1997) 337;
  P. Colangelo et al., Phys. Lett. B408 (1997) 340;
  C.A.~Dominguez et al., Nucl. Phys. B (Proc. Suppl.) 74 (1999) 313;
  K. Maltman, Phys. Lett. B462 (1999) 195;
  S. Narison, Phys. Lett. B466 (1999) 345.  

\bibitem{LatMass} 
   V. Lubicz, Nucl. Phys. B (Proc. Suppl.) 94 (2001) 116;
   R. Gupta and K.~Maltman, hep-ph/0101132.

\bibitem{BPR:98}
  J.~Prades, Nucl. Phys. B (Proc. Suppl.) 64 (1998) 253;
  J. Bijnens, J.~Prades and E.~de Rafael, Phys. Lett. B348 (1995) 226.

\bibitem{pich1} A. Pich, {\it Tau Physics: Theoretical Perspective},
  Proc. 6th Int. Workshop on Tau Lepton Physics (Victoria, Canada, 2000),
  Nucl. Phys. B (Proc. Suppl.) 98 (2001) 385;  
  {\it Tau Physics}, Proc. XIX International
 Symposium on Lepton and Photon Interactions at High Energies (Stanford, 1999),
 eds. J.~Jaros and M.~Peskin (World Scientific, Singapore, 2000) 157
 [hep-ph/9912294].

\bibitem{PDG:00} Particle Data Group, {\it Review of Particle Physics},
    Eur. Phys. J. C15 (2000) 1;
    and 2001 update [http://pdg.lbl.gov/].

\bibitem{RPS:98} G. Rodrigo, A. Pich and A. Santamaria, \pl B424 (1998)
      367.
\bibitem{ChKS:97}
   K.G. Chetyrkin, B.A. Kniehl and M. Steinhauser, \prl 79 (1997) 2184;
   W. Bernreuther and W. Wetzel, \np B197 (1982) 228;
   W. Wetzel, \np B196 (1982) 259;
   W. Bernreuther, Ann. Phys. 151 (1983) 127.

\bibitem{MW:00} K. Maltman and C.E. Wolfe, Phys. Lett. B482 (2000) 77;
  Phys. Rev. D63 (2001) 014008.  

\bibitem{EPR:91}
  G. Ecker, A. Pich and E. de Rafael, Phys. Lett. B237 (1990) 481.

\bibitem{GV:99} S. Gardner and G. Valencia, Phys. Lett. B466 (1999) 355.

\bibitem{GP:97} F. Guerrero and A. Pich, Phys. Lett. B412 (1997) 382.

\bibitem{OMNES} R. Omn\`es, {\em Nuovo Cimento} 8 (1958);\\
 N.I. Muskhelishvili, {\it Singular Integral Equations},
  Noordhoof, Groningen, 1953.

\bibitem{TR:88} T.N. Truong, Phys. Lett. B207 (1988) 495.

\bibitem{BGKO:91} M. B\"uchler et al., hep-ph/0102287, hep-ph/0102289.

\bibitem{anamar}
  M. Ciuchini et al., Nucl. Phys. B573 (2000) 201, hep-ph/0012308;
  S. Mele, \pr D59 (1999) 113011;
  F. Parodi, P. Roudeau and A. Stocchi, Nuovo Cim. A112 (1999) 833;  
A. H\"ocker, H. Lacker, S. Laplace, F. Le Diberder, hep-ph/0104062.

\bibitem{PP:95} J. Prades and A. Pich, Phys. Lett. B346 (1995) 342; J.L. Rosner, hep--ph/0011184.

\bibitem{brev}
    A.~Buras, M.~Jamin and P.H.~Weisz, \np B347 (1990) 491;
    S. Herrlich and U.~Nierste, \np B419 (1994) 292, B476 (1996) 27;
           \pr D52 (1995) 6505.

\bibitem{WO:83} L. Wolfenstein, Phys. Rev. Lett. 51 (1983) 1945.

\bibitem{RA:95} E. de Rafael, {\it Chiral Lagrangians and Kaon
   CP--Violation}, in ``CP Violation and the limits of the Standard Model''
  (TASI'94), ed. J.F.~Donoghue (World Scientific, Singapore, 1995)
  [hep-ph/9502254].

\bibitem{PER:00} 
    S. Peris and E. de Rafael, Phys. Lett. B490 (2000) 213.

\bibitem{BSW:84}
  J. Bijnens, H. Sonoda and M.B.~Wise, Phys. Rev. Lett. 53 (1984) 2367.

\bibitem{NLO_NC} M. Knecht, S. Peris and E. de Rafael, hep-ph/0102017;
Nucl. Phys. B (Proc. Suppl.) B86 (2000) 279; Phys. Lett. B457 (1999) 227.
B443 (1998) 255.

\bibitem{latka} L. Lellouch and M. L\"uscher, hep-lat/0003023;
M.F.L. Golterman and E. Pallante, JHEP 08 (2000) 023;
C.-J.D. Lin, G. Martinelli, C.T. Sachrajda and M. Testa, hep-lat/0104006.
\end{thebibliography}
\end{document}